\documentclass[iop,apj]{emulateapj}
\usepackage{color,multirow}

\usepackage[backref,breaklinks,colorlinks,citecolor=blue]{hyperref}
\usepackage[all]{hypcap}

\newcommand{\btxt}[1]{{#1}}
\newcommand{\bbtxt}[1]{{#1}}

\newcommand{\heii}{He\,{\sc ii}~$\lambda4686$}

\slugcomment{To appear in ApJ}

\shorttitle{The extraordinary H{\MakeLowercase e}\,{\sc ii}~$\lambda4686$ in $\eta$~Car}
\shortauthors{Teodoro et al.}

\usepackage{cleveref}
\crefname{equation}{equation}{equations}
\Crefname{equation}{equation}{equations}
\crefname{figure}{figure}{figures}

\begin{document}

\title{H{\MakeLowercase e}\,{\sc ii}~$\lambda$4686 emission from the massive binary system in $\eta$~C\MakeLowercase{ar}: \\ constraints to the orbital elements and the nature of the periodic minima\altaffilmark{$\ast$}\altaffilmark{$\dag$}\altaffilmark{$\ddag$}\altaffilmark{$\S$}}

\altaffiltext{$\ast$}{Based in part on observations obtained at the Southern Astrophysical Research (SOAR) telescope, which is a joint project of the Minist\'erio da Ci\^encia, Tecnologia, e Inova\c c\~ao (MCTI) da Rep\'ublica Federativa do Brasil, the U.S. National Optical Astronomy Observatory (NOAO), the University of North Carolina at Chapel Hill (UNC), and Michigan State University (MSU).}

\altaffiltext{$\dag$}{Based in part on observations made at these observatories: Pico dos Dias Observatory (OPD/LNA), Complejo Astron\'omico El Leoncito (CASLEO/CONICET), and Mt. John University Observatory (MJUO/UC).}

\altaffiltext{$\ddag$}{Based in part on observations obtained at the Cerro Tololo Inter-American Observatory, National Optical Astronomy Observatory (NOAO Prop. IDs: 2012A-0216, 2012B-0194, 2013B-0328, and 2015A-0109; PI: N. D. Richardson), which is operated by the Association of Universities for Research in Astronomy (AURA) under a cooperative agreement with the National Science Foundation and the SMARTS Consortium.}

\altaffiltext{$\S$}{Based in part on observations made with the NASA/ESA Hubble Space Telescope, obtained at the Space Telescope Science Institute, which is operated by the Association of Universities for Research in Astronomy, Inc., under NASA contract NAS 5-26555. These observations are associated with program numbers 11506, 12013, 12508, 12750, and 13054. Support for program numbers 12013, 12508, and 12750 was provided by NASA through a grant from the Space Telescope Science Institute, which is operated by the Association of Universities for Research in Astronomy, Inc., under NASA contract NAS 5-26555.}

\author{M.~Teodoro\altaffilmark{1,2}}
\affil{Astrophysics Science Division, Code 667, NASA Goddard Space Flight Center, Greenbelt, MD 20771, USA}
\altaffiltext{1}{Research Associate at Western Michigan University.}
\altaffiltext{2}{CNPq/Science without Borders Fellow.}
\email{mairan.teodoro@nasa.gov}

\author{A.~Damineli}
\affil{Instituto de Astronomia, Geof\'{\i}sica e Ci\^{e}ncias Atmosf\'{e}ricas, Universidade de S\~{a}o Paulo, R. do Mat\~{a}o 1226, Cidade Universit\'{a}ria, S\~{a}o Paulo 05508-900, Brazil}

\author{B.~Heathcote}
\affil{SASER team, 269 Domain Road, South Yarra, Victoria, 3141, Australia}

\author{N.~D.~Richardson\altaffilmark{3}, A.~F.~J.~Moffat, and L.~St-Jean}
\affil{D\'epartment de Physique, Universit\'e de Montr\'eal, CP 6128, Succursale: Centre-Ville, Montr\'eal, QC, H3C 3J7, Canada}
\altaffiltext{3}{Ritter Observatory, Department of Physics and Astronomy, The University of Toledo, Toledo, OH 43606-3390, USA}

\author{C.~Russell\altaffilmark{4}, T.~R.~Gull, and T.~I.~Madura\altaffilmark{5}}
\affil{Astrophysics Science Division, Code 667, NASA Goddard Space Flight Center, Greenbelt, MD 20771, USA}
\altaffiltext{4}{X-Ray Astrophysics Laboratory, NASA Goddard Space Flight Center, Greenbelt, MD 20771, USA}
\altaffiltext{5}{Universities Space Research Association, 7178 Columbia Gateway Dr., Columbia, MD 20146, USA}

\author{K.~R.~Pollard}
\affil{Department of Physics and Astronomy, University of Canterbury, New Zealand}

\author{F.~Walter}
\affil{Department of Physics and Astronomy, Stony Brook University, Stony Brook, NY 11794-3800, USA}

\author{A.~Coimbra and R.~Prates}
\affil{Laborat\'orio Nacional de Astrof\'{\i}ca, R. Estados Unidos, 154, Bairro das Na\c c\~oes, Itajub\'a 37504-364, Brazil}

\author{E.~Fern\'{a}ndez-Laj\'{u}s\altaffilmark{6} and R.~C.~Gamen\altaffilmark{6}}
\affil{Facultad de Ciencias Astrono\'{o}micas y Geof\'{\i}sicas, Universidad Nacional de La Plata, Paseo del Bosque s/n, La Plata, BA, \\ B1900FWA, Argentina}
\altaffiltext{6}{Instituto de Astrof\'{\i}ısica de La Plata-CONICET, Paseo del Bosque s/n, La Plata, BA, B1900FWA, Argentina}

\author{G.~Hickel and W.~Henrique}
\affil{Instituto de F\'{\i}sica \& Qu\'{\i}mica, Universidade Federal de Itajub\'{a}, Av. BPS, 1303, Pinheirinho, Itajub\'{a} 37500-062, Brazil}

\author{F.~Navarete and T.~Andrade}
\affil{Instituto de Astronomia, Geof\'{\i}sica e Ci\^{e}ncias Atmosf\'{e}ricas, Universidade de S\~{a}o Paulo, R. do Mat\~{a}o 1226, Cidade Universit\'{a}ria, S\~{a}o Paulo 05508-900, Brazil}

\author{F.~Jablonski}
\affil{Divis\~ao de Astrof\'{\i}sica, Instituto Nacional de Pesquisas Espaciais, Av. dos Astronautas, 1758, Jd. Granja 12227-010 S\~ao Jos\'e dos Campos, Brazil}

\author{P.~Luckas\altaffilmark{7}, M.~Locke\altaffilmark{8}, J.~Powles, and T.~Bohlsen}
\affil{SASER Team, 269 Domain Road, South Yarra, Victoria, 3141, Australia}
\altaffiltext{7}{The University of Western Australia, 35 Stirling Highway, Crawley WA 6009, Perth, Australia}
\altaffiltext{8}{Canterbury Astronomical Society}

\author{\btxt{R.~Chini}\altaffilmark{9}}
\affil{\btxt{Astronomisches Institut, Ruhr-Universit\"at Bochum, Universit\"atsstra\ss{}e 150, D-44780 Bochum, Germany}}
\altaffiltext{9}{Instituto de Astronom\'{\i}a, Universidad Cat\'olica del Norte, Avenida Angamos 0610, Casilla 1280, Antofagasta, Chile}

\author{M.~F.~Corcoran\altaffilmark{5} and K.~Hamaguchi\altaffilmark{10}}
\affil{CRESST and X-ray Astrophysics Laboratory, NASA Goddard Space Flight Center, Greenbelt, MD 20771, USA}
\altaffiltext{10}{University of Maryland, Baltimore County, 1000 Hilltop Circle, Baltimore, MD 21250, USA}

\author{J.~H.~Groh}
\affil{Geneva Observatory, Chemin des Maillettes 51, CH-1290 Versoix, Switzerland}

\author{D.~J.~Hillier\altaffilmark{11}}
\affil{Department of Physics and Astronomy, University of Pittsburgh, 3941 O’Hara Street, Pittsburgh, PA 15260, USA}
\altaffiltext{11}{Pittsburgh Particle Physics, Astrophysics, and Cosmology Center, University of Pittsburgh, 3941 O’Hara Street, Pittsburgh, PA 15260, USA}

\author{G.~Weigelt}
\affil{Max-Planck-Institut f\"ur Radioastronomie, Auf dem H\"ugel 69, D-53121 Bonn, Germany}

\begin{abstract}
\btxt{$\eta$ Carinae is an extremely massive binary system in which rapid spectrum variations occur near periastron. Most notably, near periastron the He\,{\sc ii}~$\lambda $4686 line increases rapidly in strength, drops to a minimum value, then increases briefly before fading away. To understand this behavior, we conducted an intense spectroscopic monitoring of the He\,{\sc ii}~$\lambda$4686 emission line across the 2014.6 periastron passage using ground- and space-based telescopes. Comparison with previous data confirmed the overall repeatability of \mbox{$EW$(\heii)}, the line radial velocities, and the timing of the minimum, though the strongest peak was systematically larger in 2014 than in 2009 by $26$\%. The \mbox{$EW$(\heii)} variations, combined with other measurements, yield an orbital period \mbox{$2022.7 \pm 0.3$ d}. The observed variability of the \mbox{$EW$(\heii)} was reproduced by a model in which the line flux primarily arises at the apex of the wind-wind collision and scales inversely with the square of the stellar separation, if we account for the excess emission as the companion star plunges into the hot inner layers of the primary's atmosphere, and including absorption from the disturbed primary wind between the source and the observer. This model constrains the orbital inclination to 135$^{\circ}$ -- 153$^{\circ}$, and the longitude of periastron to 234$^{\circ}$ -- 252$^{\circ}$. It also suggests that periastron passage occurred on \mbox{$T_{0} = 2456874.4 \pm 1.3$ d}. Our model also reproduced \mbox{$EW$(\heii)} variations from a polar view of the primary star as determined from the observed He\,{\sc ii}~$\lambda4686$ emission scattered off the Homunculus nebula.}

\end{abstract}


\section{Introduction}

Eta Carinae ($\eta$~Car) is one of the most luminous ($L_{\rm bol} \gtrsim 5 \times 10^{6}$~L$_\odot$) and most massive stars in our Galaxy \cite[e.g.][]{1997ARA&A..35....1D}. \btxt{$\eta$~Car is one of the few luminous blue variable stars, or simply LBVs \cite[][]{1978ApJ...219..445H,1984IAUS..105..233C},} with a very well constrained luminosity and age. Located at a distance of $\sim$2.3~kpc in the \btxt{very young} stellar cluster \mbox{Trumpler 16}, $\eta$~Car underwent a giant, non-terminal outburst in the early 1840s, wherein it ejected more than 10~M$_\odot$, creating the dusty, bipolar Homunculus nebula \citep{1950ApJ...111..408G,2003AJ....125.1458S,2014MNRAS.442.3316S}. The luminosity of the $\eta$~Car stellar source is derived from the enormous infrared luminosity of the surrounding Homunculus, whose dust absorbs the central stars' UV radiation and re-radiates as thermal IR radiation \citep{1997ARA&A..35....1D}.

The central source in $\eta$~Car is believed to be composed of two massive stars. On one hand, the evolutionary stage and physical parameters of the primary star are relatively well constrained: it is in the luminous blue variable (LBV) stage with a mass-loss rate of about \mbox{$10^{-3}$~M$_\odot$\,yr$^{-1}$}, a wind terminal velocity of \mbox{$420$~km\,s$^{-1}$}, and a luminosity in excess of $10^6$~L$_\odot$, which makes the primary star's spectrum dominant at wavelengths longer than $1000$~\AA~\citep{1997ARA&A..35....1D,2001ApJ...553..837H,2006ApJ...642.1098H,2012MNRAS.423.1623G}. On the other hand, due to the fact that the secondary star has never been directly observed, its physical parameters and evolutionary stage are still under debate. Nevertheless, the presence of a secondary star is inferred from the cyclic variability of the X-ray emission and changes in the ionization stage of the spectrum of the central source observed every $5.54$~yr (the so-called spectroscopic cycle or event). X-ray observations suggest that the secondary has a wind speed of $\approx3000$~km\,s$^{-1}$ and a mass-loss rate of $\sim 10^{-5}$~M$_\odot$\,yr$^{-1}$ \citep{2002A&A...383..636P}, while studies about the nebular ionization suggest that the secondary is an O-type star with $35000\lesssim T_\mathrm{eff} \lesssim 41000$~K \citep{2005ApJ...624..973V,2008MNRAS.387..564T,2010ApJ...710..729M}.

The binary nature of $\eta$~Car is very useful for constraining the current physical parameters of the stars in the system. As mentioned before, the nature of the unseen secondary star is inferred from the symbiotic-like spectrum of the system, with lines of low ionization potential (e.g. Fe\,{\sc ii}, 7.9~eV) excited by the LBV primary star and high excitation forbidden lines (e.g. [Ne\,{\sc iii}], 41~eV) attributed to photoionization by the hotter companion star. The short duration of the low excitation events \citep{2008MNRAS.384.1649D,2008MNRAS.386.2330D} and X-ray minimum \citep{2010ApJ...725.1528C} suggests a high orbital eccentricity. The first set of orbital elements, obtained from the radial velocity ($RV$) curve derived from observations of the Pa$\delta$ and P$\gamma$ lines \citep{1997NewA....2..107D}, suggested an eccentricity $e = 0.6$, orbital inclination $i\approx70\degr$, and a longitude of periastron $\omega\approx286\degr$ (note that this value refers to the orbit of the secondary in the relative orbit). In this configuration, the secondary star is `behind' the primary at periastron. \btxt{\cite{1997NewA....2..387D} pointed out that the $RV$ curve was better reproduced by adopting an orbit with higher eccentricity ($e\gtrsim0.8$) but with the same orientation as found by \citep{1997NewA....2..107D}. \cite{2001ApJ...547.1034C} showed that the first X-ray light curve observed during the 1997-8 periastron passage was well reproduced by $e = 0.9$, which was later corroborated by analysis of X-ray light curves from multiple periastron passages \cite[e.g.][]{2008MNRAS.388L..39O,2009MNRAS.394.1758P,2011ApJ...726..105P,2013PhDT.......164R}. Currently, $e=0.9$ is the value adopted by most researchers}.

There is a consensus that the $\eta$~Car binary orbital axis is closely aligned with the Homunculus polar axis at an inclination \mbox{$130\degr\lesssim i \lesssim 145\degr$} and position angle \mbox{$302\degr\lesssim PA \lesssim 327\degr$} \citep[see e.g.][]{2012MNRAS.420.2064M}. However, some residual debate exists regarding the longitude of periastron of the secondary star. On one hand, results from multi-wavelength observational monitoring campaigns, together with three-dimensional (3D) hydrodynamical and radiative transfer models of $\eta$~Car's binary colliding winds, have constrained this parameter to \mbox{$230\degr\lesssim\omega\lesssim270\degr$}, which places the primary star between the observer and the hotter companion star at periastron \citep{2007ApJ...660..669N,2007ApJ...663..522H,2008ApJ...680..705H,2008MNRAS.388L..39O,2009MNRAS.394.1758P,2009ApJ...707..693M,2010A&A...517A...9G,2010AJ....139.1534R,2011ApJ...743L...3G,2011ApJ...740...80M,2012MNRAS.420.2064M,2012ApJ...746L..18M,2012ApJ...759L...2G,2012MNRAS.423.1623G,2013ApJ...773L..16T,2013MNRAS.436.3820M,2014MNRAS.443.2475C,2015MNRAS.447.2445C,2015MNRAS.450.1388C,Richardson:2015ur}. On the other hand, there are some that favor an orientation with \mbox{$\omega=90\degr$} \citep[e.g.][]{2005MNRAS.357..895F,2007MNRAS.378..309A,2008MNRAS.390.1751K,2009MNRAS.397.1426K,Kashi:2015ul}, \btxt{which would place the companion  between the primary and the observer at periastron}.

The nature of the spectroscopic events also remains unclear. Potential scenarios include (\textit{i}) low excitation event due to blanketing of UV radiation as the secondary star plunges into the primary dense wind \citep[e.g.][]{1996ApJ...460L..49D,1998yCat..41330299D,1999ASPC..179..288D}, (\textit{ii}) an effect similar to a shell ejection \citep[e.g.][]{2002ASPC..262..267D,2003ApJ...586..432S,1984A&A...137...79Z}, (\textit{iii}) an eclipse of the secondary star by the primary's dense wind \citep{2008MNRAS.388L..39O}, and (\textit{iv}) a collapse of the colliding winds region \btxt{onto the weaker-wind secondary component} \citep{2002ecrl.confP...7G,2011ApJ...726..105P,2011ApJ...740...80M,2012ApJ...746...73T,2013MNRAS.436.3820M}. The behavior of different spectral features during periastron may actually be a result of different combinations of the above physical effects. Models that assume solely an eclipse as the origin for the spectroscopic events cannot reproduce the long duration of the minimum in the X-ray light curve, as the models predict a recovery time that is shorter than observed \citep{2009MNRAS.394.1758P}. \btxt{Moreover, the observed recovery time of the X-rays varies from cycle to cycle (\citealt{2010ApJ...725.1528C}; Corcoran et al. 2015, in prep.)}, which is almost impossible to explain in the context of a pure eclipse phenomenon. Hence a `collapse' of the colliding winds region or some similar effect that is sensitive to relatively small changes in the stellar/wind parameters of the system has been proposed in order to help explain the long duration and variable recovery of the X-rays (\citealt{2009MNRAS.394.1758P,2013MNRAS.436.3820M,2013PhDT.......164R}; Corcoran et al. 2015, in prep.). The unusual behavior of the \heii~line emission during periastron passage in $\eta$~Car is also thought to be at least partially due to a collapse of the wind-wind collision region \citep{2006ApJ...640..474M,2011ApJ...740...80M,2012ApJ...746...73T,2013MNRAS.436.3820M,2015arXiv150404940M}.

Until the discovery of a sudden increase in \heii~line intensity just before the spectroscopic event \citep{2004ApJ...612L.133S}, it was believed that $\eta$~Car had no \heii~emission. The periodic nature of the \heii~emission shows that it is directly related to $\eta$~Car's binary nature, although the exact details of the line formation mechanism are debatable \citep{2006ApJ...640..474M,2011ApJ...740...80M,2012ApJ...746...73T,2013MNRAS.436.3820M}. The large intrinsic luminosity of the \heii~line at periastron ($\sim300$~L$_\odot$) requires a luminous source of He$^+$ ionizing photons with energy greater than $54.4$~eV and/or a high flux of photons with wavelength of about $304$~\AA~(40.8~eV). In either case, it is implied that the hot companion star and/or the colliding winds play a crucial role in the \heii~line formation. The \heii~emission is likely connected to the wind-wind collision region since the post-shock primary wind is the most luminous source of photons with energies between \mbox{$54$~eV} and \mbox{$500$~eV} in the system. The short duration of the deep minimum in the \heii~emission (time interval where the line profile has completely disappeared, which lasts $\sim$1 week) and the \heii~emission's recovery and rapid fading after periastron passage \citep[see][]{2011ApJ...740...80M,2012ApJ...746...73T,2015arXiv150404940M} suggest a very compact emitting source, making it a promising probe to understand the physics involved in the periodic minima.

\begin{table*}
    \caption{Summary of the \textit{HST/STIS} observations of $\eta$~Carinae. All the observations have $R=7581$ per resolution element.\label{tab:obslog1}}
    \centering
    \begin{tabular}{ccccccc}

    \hline\hline
    \multirow{2}{*}{Program} &
    \multirow{3}{*}{P.\,I.} &
    Mapping &
    Pixel &
    \multirow{2}{*}{Slit} &
    \multirow{2}{*}{Observation} &
    \multirow{3}{*}{$SNR$\textsuperscript{a}}
    \\
    \multirow{2}{*}{ID} &
     &
    region size &
    scale &
    \multirow{2}{*}{PA} &
    \multirow{2}{*}{date} &
    
    \\
     &
     &
    (arcsec$^2$) &
    (arcsec pixel$^{-1}$) &
    \\\hline
     11506 & K.~Noll & 6.4$\times$2.0 & 0.10 & $+79.51\degr$ & 2009 Jun 30 & 955 \\ \cline{1-7}
     12508 & T.~Gull & 6.4$\times$2.0 & 0.05 & $-138.66\degr$ & 2011 Nov 20 & 272 \\ \cline{1-7}
     12750 & T.~Gull & 6.4$\times$2.0 & 0.05 & $-174.84\degr$ & 2012 Oct 18 & 368 \\ \cline{1-7}
\multirow{5}{*}{13054} & \multirow{5}{*}{T.~Gull} & \multirow{5}{*}{6.4$\times$2.0} & \multirow{5}{*}{0.05} & $-136.73\degr$ & 2013 Sep 03  & 265 \\
 &  &  &  & $-56.3\degr$ & 2014 Feb 17 & 374 \\
 &  &  &  & $+61.4\degr$ & 2014 Jun 09 & 359 \\
 &  &  &  & $+107.6\degr$ & 2014 Aug 02 & 503 \\
 &  &  &  & $+162.1\degr$ & 2014 Sep 28 & 462 \\\hline
 \multicolumn{6}{r}{Total number of spectra} & 8 \\\hline
 \multicolumn{7}{l}{\textsuperscript{a}\footnotesize{Signal-to-noise ratio per resolution element.}}\\
     
    \end{tabular}
    
\end{table*}

\begin{table*}
    \centering
    \caption{Summary of the \btxt{2014.6} ground-based observations of $\eta$~Carinae.}\label{tab:obslog2}
    \begin{tabular}{ccccc}

    \hline\hline
	\multicolumn{5}{c}{Contribution from professional observatories} \\ \hline
    Observatory & P.~I. & Telescope & Spectrograph & $N$\textsuperscript{a} \\ \hline
     \multirow{2}{*}{CTIO} & N.~Richardson & \multirow{2}{*}{$1.5$~m} & \multirow{2}{*}{{\sc chiron}} & \multirow{2}{*}{114} \\
                                        & F.~Walter &  &  & \\ \cline{1-5}
     \multirow{2}{*}{OPD} & \multirow{2}{*}{A.~Damineli} & $1.6$~m & Coud\'e & 36 \\
                                       &  & $0.6$~m & Lhires\,{\sc iii} & 57 \\ \cline{1-5}
	 SOAR & M.~Teodoro & $4.1$~m & Goodman & 37 \\
     MJUO & K.~Pollard & $1$~m & {\sc hercules} & 26 \\
     CASLEO & E.~Fern\'andez-Laj\'us & $2.15$~m & {\sc reosc dc} & 19 \\ \hline
	\multicolumn{5}{c}{Contribution from SASER members} \\ \hline
    Observer & Location & Telescope\textsuperscript{b} & Spectrograph + Camera & $N$\textsuperscript{a} \\ \hline
     P.~Luckas & Perth, Australia & $0.35$~m & Spectra L200 + Atik 314L & 17 \\
     B.~Heathcote & Melbourne, Australia & $0.28$~m & Lhires {\sc iii} + Atik 314L & 10 \\
     M.~Locke & Canterbury, New Zealand & $0.40$~m & Spectra L200 + SBIG ST-8 & 7 \\
     J.~Powles & Canberra, Australia & $0.25$~m & Spectra L200 + Atik 383L+ & 7 \\
     T.~Bohlsen & Armidale, Australia & $0.28$~m & Spectra L200 + SBIG ST-8XME & 5 \\ \hline
     \multicolumn{4}{r}{Total number of spectra} & 335 \\\hline
    \multicolumn{5}{l}{\textsuperscript{a}\footnotesize{Total number of spectra used in the present work.}}\\
    \multicolumn{5}{l}{\textsuperscript{b}\footnotesize{Except for P.~Luckas, who uses a Ritchey-Chr\'etien, the SASER team employs Schmidt-Cassegrain telescopes.}}
    \end{tabular}

\end{table*}

There have been many attempts to explain the formation and behavior of $\eta$~Car's \heii~emission \btxt{\citep[e.g.][]{2004ApJ...612L.133S,2006ApJ...640..474M,2011ApJ...740...80M,2012ApJ...746...73T,Davidson:2015gs,2015arXiv150404940M}}. The model proposed by \cite{2013MNRAS.436.3820M}, based on the results of 3D hydrodynamical simulations, presents another mechanism for explaining $\eta$~Car's \heii~emission. In this model, the \heii~emission is a result of a pseudo `bore hole' effect \citep{2010RMxAC..38...52M} wherein at phases around periastron the He$^+$ zone located deep within the primary's dense extended wind is exposed to extreme UV photons emitted from near the apex of the wind-wind collision zone. The extreme UV photons emitted around the apex of the wind-wind collision interface penetrate into the primary's He$^+$ region, producing He$^{2+}$ ions whose recombination produces the observed \heii~emission. This model is promising as it contains all of the required ingredients constrained by the observational data: a powerful source of photons with energies greater than \mbox{$54.4$~eV} (the colliding wind shocks), a relatively compact region containing a large reservoir of He$^+$ ions (the inner $3$~AU region of the dense primary wind), and a physical mechanism to explain the brief duration and timing of the observed \heii~flare around periastron (penetration of the colliding winds' region into the primary's He$^+$ core). However, this mechanism is only effective about $30$~d before through $30$~d after periastron passage, when the apex is close to or inside the He$^+$ core. Since the observations indicate that the \heii~equivalent width starts to increase about $6$ months before periastron passage, the `bore hole' effect cannot be the sole mechanism responsible for the \heii~emission; additional processes must be present before the onset of the `bore hole' effect.

In order to better understand the \heii~emission in $\eta$~Car and its relation to the binarity and the recent reports of supposed changes that might have occurred in the system \citep{2010ApJ...725.1528C,2011ApJ...740...80M}, we organized an intensive campaign to monitor the \heii~line across $\eta$~Car's 2014.6 periastron passage. Our observing campaign is the most detailed yet of an $\eta$~Car \heii~event, consisting of over 300 individual spectral observations. The main goals of the campaign were to: (1) collect data with both medium-to-high resolution \btxt{($\Delta v \lesssim 100$~km\,s$^{-1}$)} and high signal-to-noise ratio ($SNR>200$), since the line is broad and very faint at times far from the periastron events (about $-0.1$~\AA\ in equivalent width) and (2) to have daily visits for a few months around the periastron event. \btxt{The campaign was successful, generating a large database that allowed us to (1) determine the period and stability of the \heii~equivalent width ($EW$) curve, (2) constrain the orbital parameters of the system, as well as the time of periastron passages, and (3) develop a quantitative model to explain the observed variations in the \heii~equivalent width and the nature of the periodic minima.}

In the following section, we describe the observations and the data reduction and analysis. The results are presented in Section~\ref{sec:results} followed by a discussion (Section~\ref{sec:discussion}). Finally, a summary of the main results and conclusions are presented in Section~\ref{sec:conclusions}.

\section{Observations, data reduction, and analysis}\label{sec:obs}

The data presented in this work were obtained by the \mbox{$\eta$~Car International Campaign team}, composed of members  from different observatories that participated in the monitoring of the 2014.6 event of $\eta$~Car. The main characteristics of the telescopes and instruments used during the observations are summarized in \mbox{\Cref{tab:obslog1,tab:obslog2}.}

\btxt{As we had contributions from many different instruments and instrumental configurations, the $SNR$s presented in this work are given \textit{per resolution element}. The resolution element of each spectrum was measured using the mean \textit{FWHM} (full width at half maximum) obtained from a gaussian fit to a few isolated line profiles of the comparison lamp spectrum around $4741$~\AA.} 

\begin{figure}
  \centering
  \includegraphics[width=\linewidth]{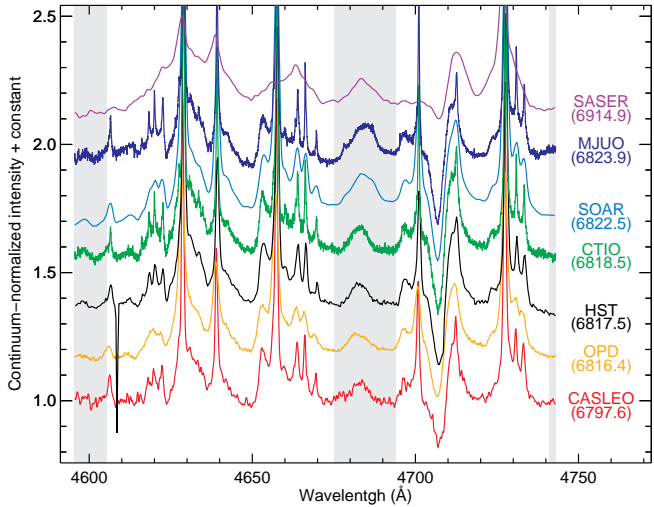}
  \caption{Typical processed and reduced spectra from collaborators of the campaign. The observatory/team and the date of acquisition (\mbox{JD-2450000}) of the spectrum are indicated on the right side of the figure. Note that, except for the SASER spectrum, which was obtained after the minimum, all of the other spectra were obtained before it and within a timeframe of about 26~d. The vertical shaded regions indicate the regions adopted for continuum (\mbox{blue: $4600\pm5$~\AA;} \mbox{red: $4742\pm1$~\AA)} and line integration \mbox{($4675$--$4694$~\AA)}.\label{fig:spec}}
\end{figure}

\btxt{The equivalent width measurements were performed homogeneously, using the protocol described in \citet{2012ApJ...746...73T}, which was adapted from \citet{2006ApJ...640..474M}. For a detailed discussion on the definition of continuum and integration regions, as well as the continuum fitting procedure, we refer the reader to those publications. For the present work, we needed to change the width of the blue continuum (see \Cref{fig:spec}) because the relatively small width previously used was susceptible to contamination by N\,{\sc II} absorption and emission components, which seemed stronger in the 2014.6 event than in the previous one \citep[see][]{Davidson:2015gs}. To dilute the influence of \bbtxt{these components} in the blue continuum region, we kept the same wavelength as before, but adopted a wider range to estimate the intensity of the continuum. Then we applied a linear fit to the blue and red continuum intensity and used the result as a baseline for the equivalent width measurements.}

\btxt{The consistency of the measurements was achieved by always measuring the equivalent width using the same method and then applying a single systematic correction to the measurements from each dataset in order to account for instrumental differences. We adopted the measurements from the integrated $2\times2$~arcsec$^2$ maps of \textit{HST/STIS} as a baseline (i.e. no corrections were applied to them) to determine the systematic correction for each observatory. The largest systematic correction used in the present work was $2$~\AA, and it was applied to the \textit{CASLEO/REOSC} dataset because of significant distortions present in the spectra due to difficulties in removing the blaze function of that spectrograph. For all the other datasets, we used systematic corrections smaller than $0.5$~\AA.}

\subsection{CTIO/CHIRON}
We monitored the system with the {\sc ctio} 1.5~m telescope and the fiber-fed {\sc chiron} spectrograph \citep[][]{2013PASP..125.1336T} from early 2012 through mid 2014 as an extension of the monitoring efforts presented by \cite{2010AJ....139.1534R,Richardson:2015ur}. The fiber is 2.7~arcsec on the sky, which is large enough so that we should not suffer from large spatial variations in the observed background nebulosity. \btxt{The resulting spectra have a resolving power between 80,000 and 100,000 and exhibit a strong blaze function}. In order to remove the blaze function and have a realistic normalized spectrum, we compared each spectrum to a spectrum obtained of HR 4468 (B9.5\,{\sc V}n), which has very few spectral features except for H$\alpha$ and H$\beta$ in the \btxt{spectral window 4500--7500~\AA}. The stability of {\sc chiron} allowed us to achieve a good rectification of the continuum without the need of frequent observations of standard stars (in fact, we only needed one for each observation mode). The wavelength calibration of the data was performed using a ThAr lamp spectrum obtained on the same night as the observations of $\eta$~Car. We also used the narrow line emission component (originating in the nebulosity around the central source whose velocity is relatively well known) to check the wavelength solution. 

\subsection{OPD}
The spectra from OPD (Observat\'orio do Pico dos Dias; operated by LNA/MCT-Brazil) were collected at the Zeiss 0.6~m telescope with the Lhires~{\sc iii} spectrograph equipped with a Atik 460EX CCD. \btxt{A set of additional spectra, with higher spectral resolution, was taken at the Coud\'e focus of the 1.6~m telescope to check for continuum normalization and heliocentric transformations of the Lhires {\sc iii} spectra.}

Data reduction was done with {\sc iraf} in the standard way. \btxt{For the Lhires~{\sc iii} dataset, the typical resolution element was about $95$~km\,s$^{-1}$, whereas for the Coud\'e spectra it was about $50$~km\,s$^{-1}$. This resolution element was enough to give information on the velocity field of the region forming the \heii~line (\textit{FWHM}$>$400~km\,s$^{-1}$). Both dataset presented a typical $SNR$ of about 550}.

In addition to $\eta$~Car, a bright \mbox{A-type} star (in general a spectrophotometric standard) was also observed in order to aid the normalization process of the stellar continuum. Wavelengths were transformed to the heliocentric reference system and checked against the narrow line components reported by \cite{1998yCat..41330299D}.

\subsection{CASLEO/REOSC}
\btxt{Spectral data of $\eta$~Car were also obtained at the Complejo Astron\'omico El Leoncito (CASLEO), Argentina, from 2014 March through August}. The frequency of observations was increased around the periastron passage during 2014 July/August.

The spectra were collected using the {\sc reosc} spectrograph in its echelle mode, attached to the 2.15-m \mbox{`J. Sahade'} telescope. A Tek $1024\times1024$~pixel$^2$ CCD (24~$\mu$m pixel), was used as detector, providing a dispersion of \mbox{$0.2$~\AA~pixel$^{-1}$}. The wavelength coverage ranges from $4200$ through $6750$~\AA.

\btxt{The normalization of $\eta$~Car spectra was performed by dividing it by the continuum of a hot star, usually $\omega$~Car or $\theta$~Car}. Additional residuals were minimized by fitting a low-order polynomial function and defining \textit{ad hoc} spectral ranges to constrain the continuum to the region \mbox{$4550$--$4750$~\AA}.

\subsection{\textit{HST/STIS}\label{sec:HST}}

High spatial sampling ($0.05$--$0.1$~arcsec per pixel) spectroscopic mapping of the region was recorded at critical binary orbital phases between 2009 June and 2014 November using the Hubble Space Telescope/Space Telescope Imaging Spectrograph (\textit{HST/STIS}; see \Cref{tab:obslog1}). The spectra of interest utilized the $52\times0.1$~arcsec$^2$ aperture with the G430M grating centered at $4706$~\AA. The mappings were accomplished by a pattern of slit positions centered on $\eta$~Car. While the first two mappings were done at a spacing of $0.1$~arcsec, subsequent mappings were done at $0.05$~arcsec spacing. Allowance for potential detector saturation was provided by a sub-array mapping directly centered on $\eta$~Car. Due to solar panel orientation constraints, the aperture position angle changed between observations. As $\eta$~Car is close to the \textit{HST} orbital pole, visits were done during continuous viewing zone opportunities, thus increasing observing efficiency more than two-fold. Spatial mappings from these data indicate that the \textit{HST/STIS} response to the central source have a \textit{FWHM}=$0.12$~arcsec.

A data cube of flux values was constructed for each spatial position in right ascension and declination at $0.05$~arcsec spacing and in velocity relative to He\,{\sc ii}~$\lambda4686$ at 25 km\,s$^{-1}$ intervals ranging from $-8,000$ to $+10,000$~km\,s$^{-1}$.

\begin{figure}
  \centering
  \includegraphics[width=\linewidth]{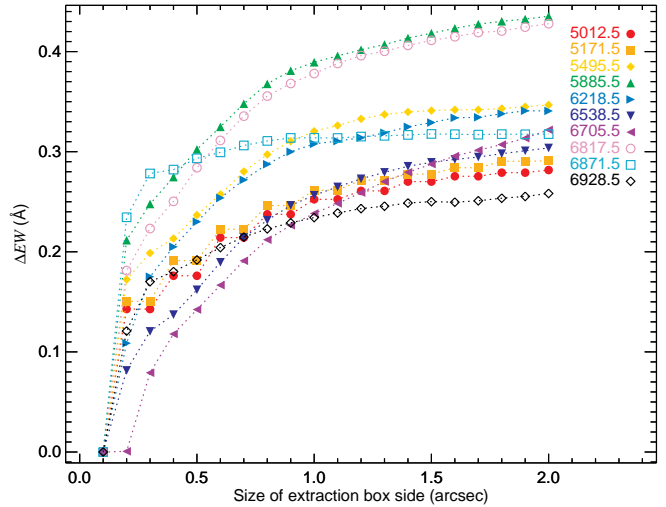}
  \caption{Aperture correction for the He\,{\sc ii}~$\lambda4686$ equivalent width measurements from \textit{HST/STIS} for each of the 10 visits (the legend indicates ${\rm JD} - 2,450,000$ for each visit). The measurements were subtracted by the equivalent width measured within a box region with dimensions $0.1\times0.1$~arcsec$^2$. The larger the extraction aperture, the larger the correction factor to be applied to the \textit{HST/STIS} measurements in order to compare with the ground-based measurements. The corrected equivalent width, \btxt{shown in the rest of this paper}, is thus given by $EW=EW^\prime+\Delta EW$, where $EW^\prime$ is the value obtained using an extraction aperture of $0.1\times0.1$~arcsec$^2$.\label{fig:apcor}}
\end{figure}

For the \textit{HST/STIS} data, the equivalent width of He\,{\sc ii}~$\lambda4686$ was measured using the same procedure as for the ground-based observations. However, unlike the space-based observations, emission from the central source cannot be separated from the surrounding nebulosity in ground-based observations due to atmospheric seeing. Hence, direct comparison between the two datasets might be hampered by unwanted contaminations, not only from nebular emission but also from continuum scattered off fossil wind structures \citep{2013ApJ...773L..16T}. For space-based observations, the contribution from such contaminations is directly proportional to the slit aperture, whereas for ground-based observations, they are always present.

\btxt{\Cref{fig:apcor} shows that, as we increase the size of the aperture, the amount of He\,{\sc ii}~$\lambda4686$ emission relative to the continuum decreases. This indicates that the He\,{\sc ii}~$\lambda4686$ emission and adjacent continuum do not come from the same volume}. Hence, in order to properly compare the \textit{HST/STIS} measurements with those obtained by ground-based telescopes, we measured the He\,{\sc ii}~$\lambda4686$ equivalent width using the final spectrum obtained from summing up the spectra from the entire $2\times2$~arcsec$^2$ mapping region of the \textit{HST/STIS} data cube.

\begin{figure*}
  \centering
  \includegraphics[width=\linewidth]{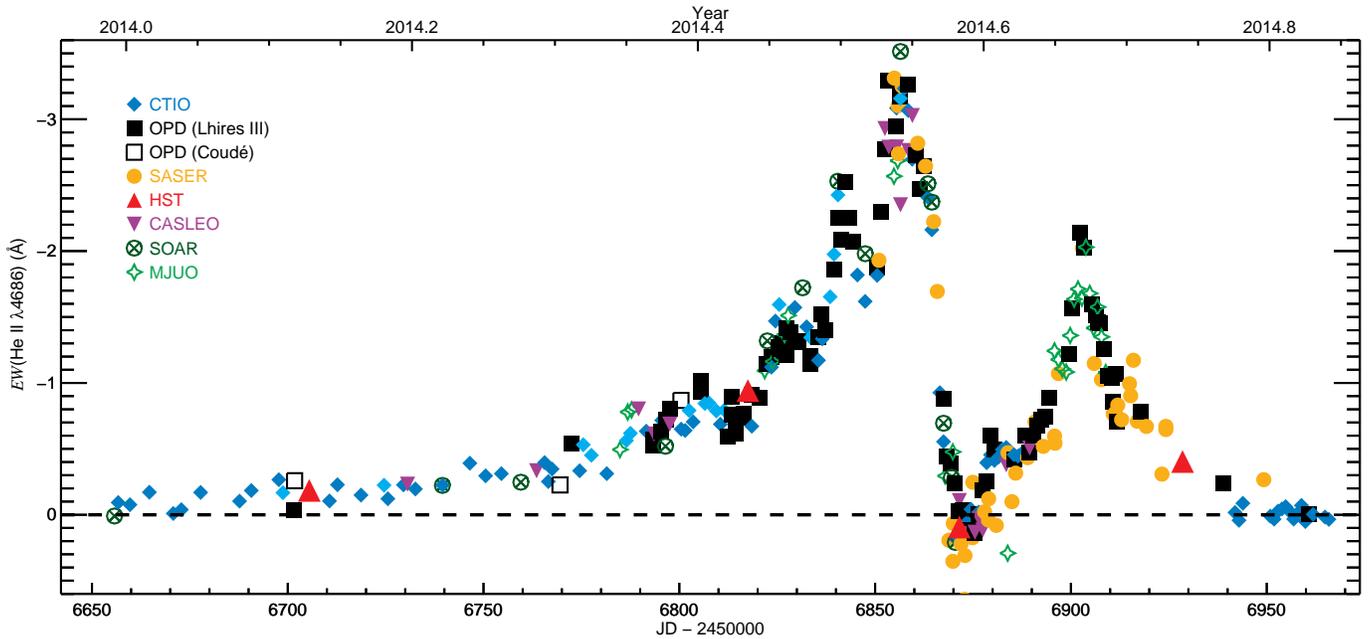}
  \caption{Equivalent width of He\,{\sc ii}~$\lambda4686$. The different symbols and colors identify the contributions from each listed observatory/team. Note that the \textit{HST/STIS} measurements were obtained using the integrated spectrum within a $2\times2$~arcsec$^2$ region in order to account for the aperture correction required to adequately compare the space- and ground-based measurements. \btxt{A table containing all the measurements and information about each spectrum is available for download online (see Appendix).}\label{fig:ew}}
\end{figure*}

\subsection{SOAR/Goodman}
The data obtained with the Goodman spectrograph were processed and reduced using standard {\sc iraf} tasks to correct them for bias and flat-field, as well as to perform the extraction and wavelength calibration of the spectra. For the latter, we used a CuAr lamp to determine a low order (between 3 and 5) Chebyshev polynomial solution for the pixel-wavelength correlation. Observations of a hot standard star (HD~303308; O4 {\sc v}) were obtained -- either just before or following those of $\eta$~Car -- in order to correct the spectra by the low frequency distortions caused by instrumental response. The final product was a dataset of spectra with a $SNR$ per resolution element typically in the range from 200 to 1000 (90\% of the data) and spectral resolution element of about $85$~km\,s$^{-1}$.

\subsection{MJUO/Hercules}

\begin{table}

    \centering
    \caption{\btxt{Additional data from the 2009.0 event obtained by R.~Chini using the BESO spectrograph on the Hexapod telescope}.}\label{tab:beso}
    \begin{tabular}{ccc}

    \hline\hline
    JD & \textit{EW} (\AA) & V\textsuperscript{a} (km\,s$^{-1}$) \\\hline
    2454846.0 & $-0.86\pm0.10$ & $-237.7$ \\
    2454851.8 & $-0.17\pm0.10$ & $-46.7$ \\
    2454871.8 & $-0.60\pm0.10$ & $+1.6$ \\
	2454874.8 & $-0.86\pm0.10$ & $-91.6$ \\
	2454877.8 & $-0.88\pm0.10$ & $-87.4$ \\
	2454882.8 & $-1.21\pm0.10$ & $-92.9$ \\
	2454887.8 & $-0.97\pm0.10$ & $-122.7$ \\
	2454891.8 & $-0.67\pm0.10$ & $-56.2$ \\
	2454896.8 & $-0.63\pm0.10$ & $-56.2$ \\ \hline
     \multicolumn{3}{l}{\textsuperscript{a}\footnotesize{Velocity of the peak.}}\\
    \end{tabular}

\end{table}

\btxt{Observations for the campaign were taken during 2014 May through 2014 October.}  The reduction software used was an in-house sequence of packages written using {\sc matlab}. We typically obtained one to three spectra of $\eta$~Car with exposure time varying between 600 and 1200~s, and one spectrum of the bright hot star $\theta$~Car, with exposure time between 300 and 600~s, depending on sky conditions. The hot star was observed in order to determine the continuum and telluric features for the normalization process. We took a Th-Ar lamp spectrum before and after each science exposure for precise wavelength calibration. Flat-fields were taken at the beginning or end of each night in order to correct the science data for variations on the detector response.

By tracing the orders in the Th-Ar spectral images along the axes defined by the flat fields and using the location of the spectral lines used for wavelength calibration, a full set of wavelength-calibrated axes is obtained along which the stellar orders can be traced. Median filtering was used to remove cosmic rays. \btxt{The spectral orders of each science image were extracted and merged into a single 1D spectrum}.

\subsection{SASER}
All the data obtained by the members of the Southern Astro Spectroscopy Email Ring (\href{http://saser.wholemeal.co.nz}{SASER\footnote{\url{http://saser.wholemeal.co.nz}}}) were fully processed and reduced by them using the standard basic procedure for long slit observations, namely, correction for the bias level and pixel-to-pixel variations, extraction of the spectrum, and wavelength calibration using a comparison lamp. Most of the data were delivered without any continuum normalization. \btxt{However, they presented variations in intensity within the range $4600$--$4750$~\AA~that was accounted for by using a linear fit to this region.}

\subsection{Additional data from the 2009.0 event obtained with HPT/BESO}
\btxt{In this paper, we used 9 additional high resolution ($R \sim 50000$) spectra covering the wavelength range from 3620 to 8530~\AA~to study the variations in the He\,{\sc ii}~$\lambda4686$ strength during the recovery and fading phases after periastron passage. These additional data were obtained with the Bochum Echelle Spectrograph for the Optical \cite[BESO;][]{2011MNRAS.411.2311F} attached to the 1.5\,m Hexapod-Telescope (HPT) at the Universit\"atssternwarte Bochum on a side-hill of Cerro Armazones in Chile. All spectra were reduced with a pipeline based on a MIDAS package adapted from FEROS, the \bbtxt{similar} ESO spectrograph on La Silla. The results of our measurements for these additional data are listed on \Cref{tab:beso}.}

\section{Results}\label{sec:results}

The high data quality and frequency of observations allowed us to analyze and characterize the variations of the \heii~emission line in unprecedented detail. \Cref{fig:ew} shows the \heii~equivalent width measurements, \textit{EW}(\heii), for the entire 2014.6 campaign.

\subsection{The period of the spectroscopic cycle derived from the {\rm He}\,{\sc ii}~$\lambda4686$ monitoring}\label{sec:period}
\btxt{To determine the period of the spectroscopic cycle for He\,{\sc ii}~$\lambda4686$, we analyzed the equivalent width measurements using two approaches. (1) We used data encompassing only the last three periastron passages, and then only the data from the interval of sharp decrease in equivalent width prior to the deep minimum. (2) We used the entire equivalent width curve and all datasets, which includes the 1992.4, 2003.5, 2009.0, and 2014.6 events}

First, we focused our attention on the decreasing phase of the equivalent width\footnote{\btxt{Despite the convention of negative values for equivalent width of emission lines (which we kept in the presentation of the data), throughout this paper we will be talking about the variations in equivalent width in terms of its absolute value.}} just before the onset of the minimum. During this phase, the equivalent width rapidly decreases from its absolute maximum to a minimum level, which seems to be close to zero. One of the difficulties in assessing the real minimum intensity level is that we need data with both high resolving power and high signal-to-noise, which is not always available for a long-term dedicated monitoring. Also, small variations in the spectrum induced by data processing, reduction, and/or normalization of the spectra are the major contributors to stochastic fluctuations during the minimum intensity phase.

\begin{figure}
  \centering
  \includegraphics[width=\linewidth]{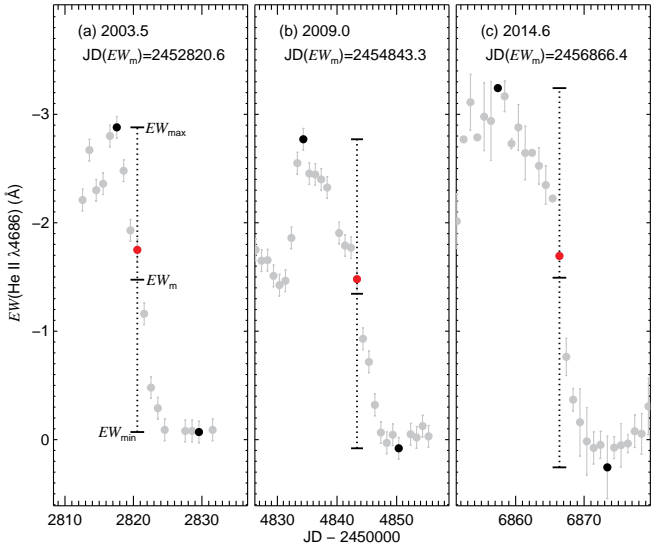}
  \caption{Period determination using the equivalent width of He\,{\sc ii}~$\lambda4686$ around periastron passage for the last 3 events: (a) 2003.5, (b) 2009.0, and (c) 2014.6. The two black dots in each panel indicate the observed maximum and minimum \mbox{$EW$(\heii)}, from where the mean equivalent width, \mbox{$EW_\mathrm{m}$}, was determined. The red dot indicates the observation nearest to \mbox{$EW_\mathrm{m}$}, whereas the vertical dotted line indicates the corresponding time of the observation, \mbox{JD($EW_\mathrm{m}$)}, indicated on the top of each panel. The period was then determined as the mean of the differences $2014.6-2009.0$ and $2009.0-2003.5$. The mean period of the He\,{\sc ii}~$\lambda4686$ obtained by this method was $2022.9\pm0.2$~d.\label{fig:period}}
\end{figure}

We noted that, as opposed to 2003.5 and 2009.0, the decreasing phase for the 2014.6 periastron passage did not occur at a linear rate. Therefore, instead of adopting the methodology used by \citet{2008MNRAS.384.1649D} for the disappearance of the narrow component of the He\,{\sc i}~$\lambda6678$ line, we used the approach suggested by \cite{2011ApJ...740...80M}. This method consists in finding the minimum ($EW_\mathrm{min}$) and maximum ($EW_\mathrm{max}$) value for the equivalent width during periastron passage, ignoring the time when they occur. Then, the mid-point is determined by $EW_\mathrm{m}=(EW_\mathrm{min}+EW_\mathrm{max})/2$. Next, the \textit{observed} equivalent width that is nearest to $EW_\mathrm{m}$ is found, for which the corresponding time, JD($EW_\mathrm{m}$), is taken as reference. Applying this procedure to consecutive periastron passages allows us to determine the period by calculating the difference between JD($EW_\mathrm{m}$)'s. \Cref{fig:period} illustrates this methodology and shows the results for data from the last three events (2003.5, 2009.0, and 2014.6). The mean period obtained by using this approach was $2022.9\pm0.2$~d, where the uncertainty is the standard deviation of the mean (the standard error is $0.14$~d).

\btxt{We also used the entire curve (including all the observations from 1992.4 up to 2014.6) to determine the period for the He\,{\sc ii}~$\lambda4686$. We folded the equivalent width curve} using trial periods to determine which period would result in the least dispersion of the data. This method, called phase dispersion minimization \citep[PDM;][]{1978ApJ...224..953S}, is frequently used to search for periodic signals in the light curve of eclipsing binary systems. In this work, we adopted the Plavchan algorithm \citep[][]{2008ApJS..175..191P}, which is a variant of the PDM method. The Plavchan algorithm folds the light curve to trial periods and, for each period, computes the $\chi^2$ difference between the original and the box-car smoothed data, but only for a predefined number of worst-fit subset of the data. Thus, the best period is the one that produces the lowest $\chi^2$ value.

\begin{figure}
  \centering
  \includegraphics[width=\linewidth]{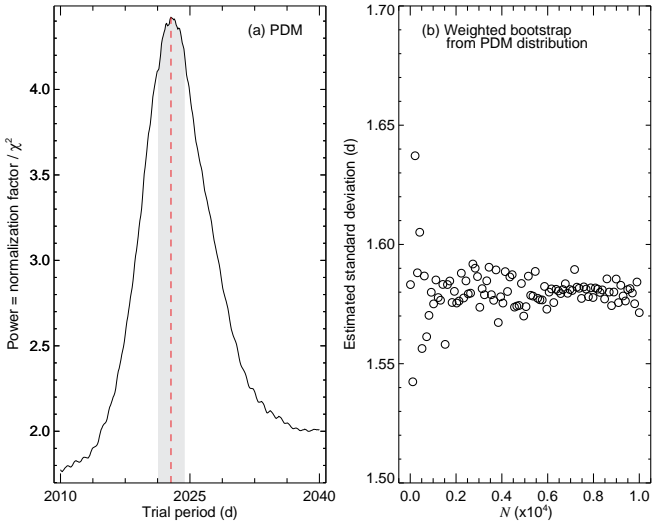}
  \caption{(a) He\,{\sc ii}~$\lambda4686$ period determination using phase dispersion minimization analysis (PDM) of the entire dataset covering the \btxt{5} last events (1992.4, \btxt{1998.0,} 2003.5, 2009.0, and 2014.6). The analysis was performed using the Plavchan period-finding algorithm, resulting in a period of $2022.8$~d. The standard deviation of the mean (shaded area in (a)) was estimated by using a weighted bootstrap technique, suggesting a value of $\pm1.58$~d.\label{fig:period2}}
\end{figure}

Using the \href{http://exoplanetarchive.ipac.caltech.edu/cgi-bin/Periodogram/nph-simpleupload}{Periodogram Service} available at the \href{http://exoplanetarchive.ipac.caltech.edu/index.html}{NASA Exoplanet Archive}\footnote{\url{http://exoplanetarchive.ipac.caltech.edu}.}, we tested a sample of 300 trial periods, equally distributed between 2010 and 2040~d. The number of elements of the worst-fit subset was set to 50 and we adopted a smoothing box-car size of 0.01 in phase. With these parameters, the result of the analysis using the Plavchan algorithm is shown in \Cref{fig:period2}a.

\begin{table}
    \centering
    \caption{Period of the spectroscopic cycle determined by different methods (adapted from \protect\citealt{2008MNRAS.384.1649D}).}\label{tab:period}
    \begin{tabular}{cc}

    \hline\hline
    Method & Period\textsuperscript{a} (d) \\\hline
    $V$ band & $2021\pm2$ \\
    $J$ band & $2023\pm1$ \\
    $H$ band & $2023\pm1$ \\
    $K$ band & $2023\pm1$ \\
    $L$ band & $2023\pm2$ \\
    Fe\,{\sc ii}~$\lambda6455$ P~Cyg radial velocity & $2022\pm2$ \\
    He\,{\sc i}~$\lambda6678$ broad comp. radial velocity & $2022\pm1$ \\
    Si\,{\sc ii}~$\lambda6347$ EW & $2022\pm1$ \\
    Fe\,{\sc ii}~$\lambda6455$ P~Cyg EW & $2021\pm2$ \\
    He\,{\sc i}~$\lambda6678$ narrow comp. EW & $2026\pm2$ \\
    He\,{\sc i}~$\lambda10830$ EW & $2022\pm1$ \\
    X-ray light curve (1992--2014) & $2023.5\pm0.7$ \\
    He\,{\sc ii}~$\lambda4686$ EW (1992--2014) & $2022.9\pm1.6$ \\\hline
    Weighted mean $\pm$ standard error & $2022.7\pm0.3$ \\
    \hline
    \multicolumn{2}{l}{\textsuperscript{a}\footnotesize{The error is the standard deviation of the mean.}}\\
    \end{tabular}

\end{table}

The maximum of the PDM power distribution, which corresponds to the minimum $\chi^2$, occurred for a period of $2022.8$~d. Although the uncertainty associated with this result cannot be obtained directly from the PDM analysis, an estimate of the standard deviation of the mean can be determined by using a weighted bootstrap technique with resampling, where a random sample with a predefined size is drawn from the original period dataset (in this case, within the range $2010$--$2040$~d). The probability of drawing a given period was determined by the PDM distribution itself and we restricted the size of each drawn sample to be 75\%~the size of the original dataset. Repeating this procedure $N$ times, where \mbox{$N\rightarrow \infty$}, allows us to obtain an estimate of the true standard deviation of the mean. \btxt{\Cref{fig:period2}b shows that this method rapidly converges: for $N \gtrsim 5000$ the dispersion of the results is about 1\%, suggesting a standard deviation of the mean of about $\pm1.6$~d}.

\btxt{The results from the two methods described in this section are consistent and can be used to obtain a mean period of $2022.9\pm1.6$~d}.

\subsection{Reassessing the mean period of the spectroscopic cycle and its uncertainty}

As an update to the previous work by \citet{2008MNRAS.384.1649D}, we used the new result from \heii~in combination with previous period determinations in order to obtain a mean period of the spectroscopic cycle. \Cref{tab:period} lists some of the methods used to determine the period. That table is an adapted version of table~2 from \citet{2008MNRAS.384.1649D}, which now includes an updated value for the period determined from the X-ray light curves, including data from 1992--2014 (Corcoran et al., in preparation), and the new determination from the He\,{\sc ii}~$\lambda4686$ equivalent width (from 1992--2014).

The weighted mean period was determined by a two-step approach. First, an unweighted mean of the 13 periods listed in \Cref{tab:period} was determined. Then, a new mean was determined by weighting each period by its absolute difference regarding the unweighted mean. This approach has the benefit of not being so sensitive to discrepant measurements, which has a great impact on the uncertainties in the period determination.

\begin{figure}
  \centering
  \includegraphics[width=\linewidth]{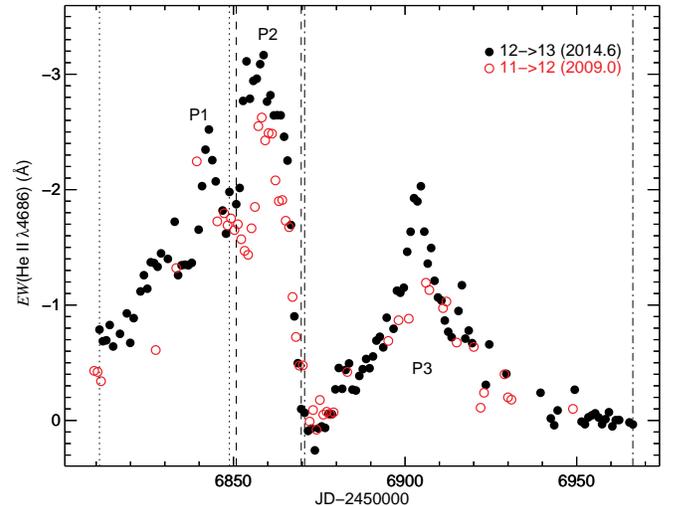}
  \caption{Comparison between the equivalent width of He\,{\sc ii}~$\lambda4686$ around periastron passage for the 2014.6 event (black filled circles) and that of 2009.0, shifted by 2022.8~d (red open circles). The vertical lines delimit the time interval used to check for variability in P1 (dotted lines), P2 (dashed lines), and P3 (dash-dotted lines) separately.\label{fig:stat}}
\end{figure}

The result suggests a weighted mean period for the spectroscopic cycle of \mbox{$2022.7\pm0.3$~d}, where the error is the standard error of the mean (the standard deviation is $0.9$~d). Thus, the mean period has not changed with the new measurements, but its uncertainty has considerably been reduced. Evidently, this improvement was mainly due to the fact that we adopted a weighted approach to the period determination, and not because we included an updated measurement from X-rays and a new datum from \heii.

\subsection{\btxt{The recurrence of the EW curve}}
We tested the hypothesis that the overall distribution of the equivalent width of the He\,{\sc ii}~$\lambda4686$ around the event is recurrent, i.e., it comes from the same distribution. We limited our analysis to the past two events (2009.0 and 2014.6) because they were monitored over almost one year (centered on the event) at a \btxt{relatively} high time sampling.

The period obtained from the He\,{\sc ii}~$\lambda4686$ equivalent width was used to fold the \btxt{observed curve} of the 2009.0 event by \btxt{2022.9}~d in order to compare it with the recent 2014.6 event. \btxt{Three non-parametric statistical tests were employed to address the similarities between the \btxt{equivalent width curves} from 2009.0 and 2014.6: the Anderson-Darling $k$-sample test \citep{Scholz:1987eq,Knuth:2011fq}, the two-sample Wilcoxon test \citep[also known as Mann-Whitney test;][]{Hollander:1973wz,Bauer:2012ef}, and the Kolmogorov-Smirnov test \citep{Birnbaum:1951fy,Conover:1971wa,Durbin:1973tp,Marsaglia:2003vb}. We used the {\sc R} package \cite[][]{Team:2014wf} to perform the statistical analysis.}

All of the statistical tests used here are non-parametric, which means that they make no assumption on the intrinsic distribution of the samples under comparison. The null hypothesis, $H_0$, for these tests is that the samples come from the same parent distribution. The tests return a parameter, called test \btxt{statistic} ($T_{\rm obs}$), which is then compared with a critical value ($T_{\rm c}$) that is calculated based on the size of the samples and on the chosen significance level, $\alpha$. Thus, the null hypothesis is rejected at the significance level $\alpha$ if \mbox{$T_{\rm obs} \geq T_{\rm c}$}, which is known as the traditional method. Another equally valid approach is to calculate the probability, under the null hypothesis, of getting a $T$ as large as $T_{\rm obs}$, which is called the $p$-value method. In this case, the $p$-value must be compared to the chosen significance level, $\alpha$, and the null hypothesis is rejected when \mbox{$p(T\geq T_{\rm c}) \leq \alpha$}. For the analysis presented here, we established a significance level of $\alpha=0.05$, and we always discard $H_0$ whenever one of the two methods rejects the null hypothesis.

\Cref{fig:stat} shows the \btxt{equivalent width curve} (ranging from \mbox{JD = 2454787.5} through 2454926.0 for 2009.0 and from \mbox{JD = 2456810.96} through 2456966.37 for the 2014.6 \btxt{curve}) for which we performed these statistical \btxt{analyses}. \Cref{tab:stat} indicates that both conditions, \mbox{$T_{\rm obs} \leq T_{\rm c}$} and \mbox{$p(T\geq T_{\rm c}) \geq \alpha$}, are true for all tests performed on the \btxt{entire curve}, suggesting that we cannot reject the null hypothesis. Note that, under the assumption that the null hypothesis is valid, a value of \mbox{$p(T\geq T_{\rm c}) \geq 0.1$} is classified as \textit{weak evidence} against the null hypothesis at the significance level $\alpha=0.05$, whereas \mbox{$p(T\geq T_{\rm c}) \geq 0.2$} is considered as \textit{no evidence} against the null hypothesis \citep{Fisher:1925vv,Fisher:1935uc}. Thus, the results of the analysis for the \btxt{entire curve} suggest that there is only weak evidence of significant changes in the \heii~equivalent width from one cycle to another.

\begin{table}
    \centering
    \caption{\btxt{Results of the two-sample statistical analysis, performed on the equivalent width curve of 2009.0 and 2014.6 (\Cref{fig:stat}), to test for the null hypothesis ($H_0$) that the two curves have the same distribution. $T_{\rm obs}$ is the test statistic and \mbox{$p(T\geq |T_{\rm obs}|)$} is the probability, assuming that $H_0$ is valid, of obtaining a test statistic as extreme as observed. $N_1$ and $N_2$ are the sample size for 2009.0 and 2014.6, respectively.}}\label{tab:stat}
    \begin{tabular}{ccccc}

    \hline\hline
    Hypothesis test & $|T_{\rm obs}|$ & $T_{\rm c}$\textsuperscript{a} & \mbox{$p(T\geq |T_{\rm obs}|)$} & Reject  $H_0$?\\\hline
     \multicolumn{5}{c}{Entire curve ($N_1=54$ and $N_2=122$)} \\\hline
     AD\textsuperscript{b} & 0.156 & 1.960 & 0.300 & No \\
     Wrs\textsuperscript{c} & 0.626 & 1.960 & 0.532 & No \\
     KS\textsuperscript{d} & 0.127 & 0.222  & 0.583 & No \\\hline
     \multicolumn{5}{c}{P1 only ($N_1=11$ and $N_2=32$)} \\\hline
     AD & 1.041 & 1.960 & 0.120 & No \\
     Wrs & 0.334 & 1.960 & 0.738 & No \\
     KS & 0.364 & 0.475 & 0.229 & No \\\hline
     \multicolumn{5}{c}{P2 only ($N_1=20$ and $N_2=20$)} \\\hline
     AD & 6.195 & 1.960 & 0.001 & Yes \\
     Wrs & 3.00 & 1.960 & 0.003 & Yes \\
     KS & 0.600 & 0.430 & 0.002 & Yes \\\hline
     \multicolumn{5}{c}{P3 only ($N_1=23$ and $N_2=70$)} \\\hline
     AD & -0.021 & 1.960 & 0.365 & No \\
     Wrs & 0.045 & 1.960 & 0.964 & No \\
     KS & 0.199 & 0.327 & 0.501 & No \\\hline\hline
    \multicolumn{5}{l}{\textsuperscript{a}\footnotesize{Critical value of the test for a significance level $\alpha=0.05$.}}\\
    \multicolumn{5}{l}{\textsuperscript{b}\footnotesize{AD: Anderson-Darling test.}}\\
    \multicolumn{5}{l}{\textsuperscript{c}\footnotesize{Wrs: Wilcoxon rank sum test.}}\\
    \multicolumn{5}{l}{\textsuperscript{d}\footnotesize{KS: Kolmogorov-Smirnov test.}}\\
    \end{tabular}

\end{table}

We also looked for changes in the intensity of each peak (P1, P2, and P3) separately. For this, we defined three data subsets, each one containing one peak, to be analyzed in the same way as the \btxt{entire curve}. The first subset, containing P1, is composed of data within the range 2454787.5--2454826.36 for 2009.0 and within 2456810.96--2456848.81 for 2014.6. The second subset, which contains P2, is composed of data within 2454827.36--2454846.29 for 2009.0 and within 2456850.81--2456869.74 for 2014.6. Finally, the third subset contains P3 and is composed of data within 2454847.29--2454926 for 2009.0 and within 2456870.73--2456966.37 for 2014.6.

The results shown in \Cref{tab:stat} suggest that P1 and P3 did not change significantly between 2009.0 and 2014.6. In contrast, P2 had a statistically significant increase in \btxt{strength} over this period\btxt{; the mean difference between the two epochs is about 0.53~\AA, which corresponds to a relative increase of about 26\%~from 2009.0 to 2014.6.}

Caution is advised regarding the lack of variability of P1 and P3, since their different sample size (resulting from different frequencies of observations) might have some influence on the statistical analysis in the case where significant changes occurred during epochs not covered by the monitoring. \bbtxt{Since P2 does not suffer from different sampling between 2009.0 and 2014.6, the result obtained for this peak is more reliable than that for P1 and P3. Nevertheless, even if we adopt a different time interval to be used for the statistical analyses of each peak (e.g. reducing the time interval for P3 to the range 2456880--2456935), the outcome of the tests remain unchanged. Therefore, our results suggest that P3 might be composed of a broad and a narrow component. The former has been detected in previous events, but the latter was only detected in 2014.6 due to the better time sampling of observations. Further discussion on this discrepancy in P3 between 2009.0 and 2014.6 is presented in Section \ref{sec:remarks}.}

\subsubsection{Comparing equivalent width measurements from HST/STIS and ground-based}

There has been a debate about discrepancies resulting from measuring the equivalent width of the He\,{\sc ii}~$\lambda4686$ using \textit{HST/STIS} and ground-based observations. Recently, \citet{Davidson:2015gs}, based on five measurements using \textit{HST/STIS} data, suggested that the He\,{\sc ii}~$\lambda4686$ equivalent width for the 2014.6 event was systematically different from the past cycles. Their conclusion relies on the direct comparison between \textit{HST/STIS} and ground-based observations for the past two cycles (2009.0 and 2014.6).

In the case of $\eta$~Car, \textit{HST/STIS} observations have undoubtedly higher quality than the ground-based ones in the sense that they allow us to obtain the spectrum of the central source with relatively less contamination from the surrounding nebulosities. Nevertheless, space-based observations could not be scheduled so as to have the same time coverage possible from the ground, which were performed almost on a daily basis (at least around the event, for the past two cycles). Also, ground-based optical spectra frequently are obtained with a much higher resolving power than possible with \textit{HST/STIS}, allowing us to have better understanding of the line morphologies.

\begin{figure}
  \centering
	\includegraphics[width=\linewidth]{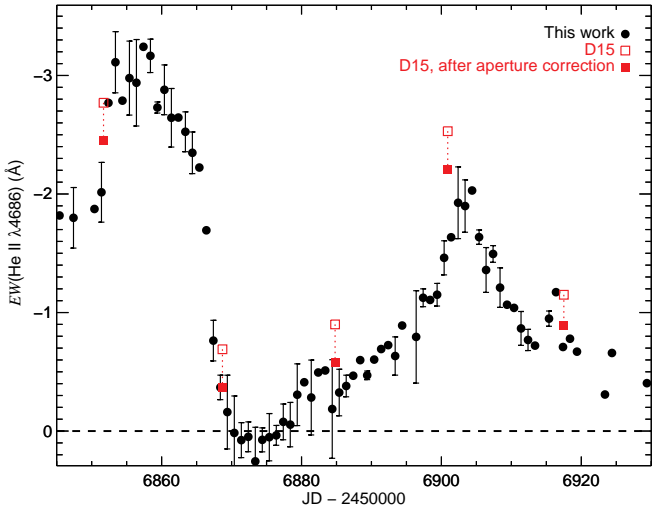}
    \caption{Comparison between the measurements presented in this work \btxt{for the 2014.6 event} (using ground- and space-based telescopes) and those from \protect\citet[][D15, using \textit{HST/STIS} observations]{Davidson:2015gs}, before (red open squares) and after (red solid squares) applying aperture corrections to their measurements.\label{fig:davidson}}
\end{figure}

Thus, it is a good practice to compare the data obtained with \textit{HST/STIS} with those obtained from ground-based telescopes. However, for the reasons mentioned in \Cref{sec:HST}, aperture corrections must be applied to the measurements obtained from space-based telescopes in order to be properly compared with the ground-based measurements. This correction must be performed by summing up the observed equivalent width and the aperture correction factor obtained from \Cref{fig:apcor}. \Cref{fig:davidson} shows the result of performing such corrections to the measurements published in \citet{Davidson:2015gs}. The amount of correction for each measurement was determined from our \textit{HST/STIS} data using the closest observation. After the corrections were applied, no significant differences are observed (within the uncertainties of the measurements).

Although we did not detect significant variations in the overall behavior of the \btxt{equivalent width curve} of the He\,{\sc ii}~$\lambda4686$ equivalent width in a timeframe of 5.54~yr, we \btxt{cannot discard the possibility of changes over longer timescales}. However, this can only be adequately addressed by continuously monitoring $\eta$~Car over several more cycles.

\subsection{Detection of {\rm He}\,{\sc ii}~$\lambda4686$ emission around apastron}

\begin{figure}
  \centering
	\includegraphics[width=\linewidth]{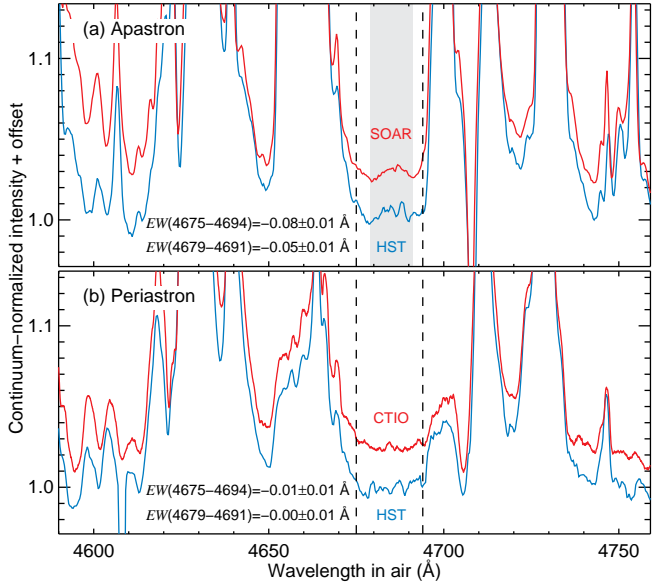}
    \caption{Ground- and space-based spectra taken at phases around (a) apastron and (b) periastron, showing the consistency of the equivalent width ($EW$) measured in both data. For clarity, the ground-based spectra were shifted vertically by an offset of $0.025$. The vertical dashed lines indicate the wavelength interval from 4675--4694~\AA, usually adopted for the equivalent width determination, whereas the shaded region indicates the range 4679--4691~\AA, which was adopted at apastron in order to minimize the contribution from the wings of the neighboring iron lines to the flux within the integration region. The $EW$ indicated in each panel is the minimum value for each wavelength interval. See main text for more details.\label{fig:emissionatapastron}}
\end{figure}

A comparison between the He\,{\sc ii}~$\lambda4686$ line profile at phases around apastron and periastron is shown in \Cref{fig:emissionatapastron}. That figure leaves no doubt about the positive detection of this line in emission at phases around apastron for the last cycle (between 2009.0 and 2014.6). Indeed, as can be seen from \Cref{fig:emissionatapastron}a, there is a noticeable hump in the range from 4675 to 4694~\AA, where the He\,{\sc ii}~$\lambda4686$ is located at phases around periastron. As an illustration of what a zero He\,{\sc ii}~$\lambda4686$ emission would look like, \Cref{fig:emissionatapastron}b shows two spectra taken at the onset of the \heii~deep minimum (2014 Jul 31), using CTIO/CHIRON and \textit{HST/STIS} data. The region where the He\,{\sc ii}~$\lambda4686$ line was previously detected is flat, showing no evident line profile.

Due to the restricted wavelength interval adopted to calculate the \heii~equivalent width \mbox{(4675--4694~\AA)}, contamination from the broad component of the iron emission lines on both sides of the integration region are always included in the measurements at all phases. This effect can be especially significant at phases around apastron, when the intensity of the \heii~line is relatively low and contaminations become stronger.

Although we cannot reliably determine the exact magnitude of the contaminations, we can estimate a minimum value for the \heii~equivalent width at apastron by reducing the size of the integration region so that we keep only the observed line profile. As can be seen in \Cref{fig:emissionatapastron}a, the \heii~line profile is evident in the range \mbox{4679--4691~\AA} (shaded area in that figure). The equivalent width measured within this reduced region, at apastron, was about $-0.051$~\AA~for SOAR and $-0.048$~\AA~for \textit{HST/STIS}, resulting in a minimum of $-0.05\pm0.01$~\AA. For the sake of completeness, the equivalent width measured within the wider wavelength interval (4675--4694~\AA) was $-0.083$~\AA~for SOAR and $-0.075$~\AA~for \textit{HST/STIS}, which implies an equivalent width of $-0.08\pm0.01$~\AA. Hence, the minimum equivalent width of the \heii~emission line at apastron is $-0.07\pm0.01$~\AA.

\section{Discussion}\label{sec:discussion}

\begin{figure}
  \centering
  \includegraphics[width=\linewidth]{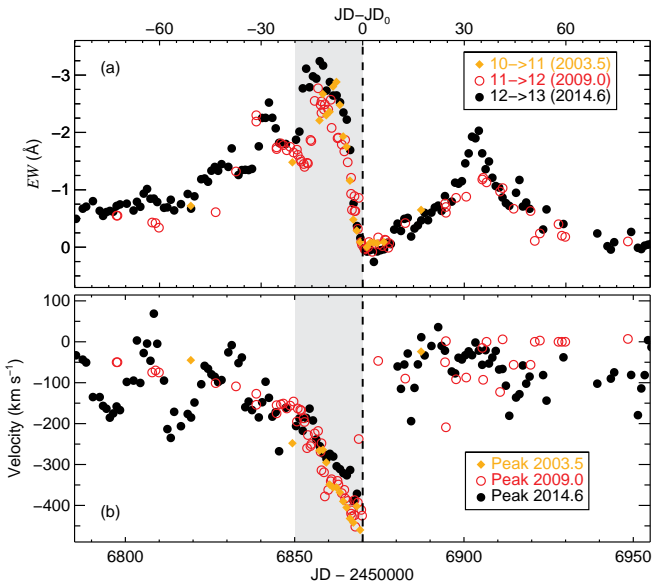}
  \caption{(a) He\,{\sc ii}~$\lambda4686$ \btxt{equivalent width curve} for the last three events folded by \btxt{$2022.9$}~d. (b) Doppler velocity measurements. The shaded region indicates the period where the peak  velocity rapidly changes to more blue-shifted velocities, which repeated for the past three events. This kinematic behavior seems to be related only with P2.\label{fig:vel}}
\end{figure}

The phase-locked behavior of the overall He\,{\sc ii}~$\lambda4686$ \btxt{equivalent width curve} (including P1, P2, and P3; see \Cref{fig:vel}a) suggests that at least the bulk of the line strength is due to non-stochastic processes occurring at phases close to periastron passages. However, due to intricate changes in the line profile, it is not clear yet where such a large amount of He\,{\sc ii}~$\lambda4686$ emission is located, how extended it is, nor how the line emission mechanism changes with time. Nevertheless, our results have the potential to shed light on the dominant mechanism behind the changes in the \heii~\btxt{equivalent width curve}.

There is a consensus that, at least during periastron passages, \heii~emission should be produced close to the WWC region, most likely in the dense, cool, pre-shock primary wind \citep{2006ApJ...640..474M,2012ApJ...746...73T,2013MNRAS.436.3820M}. This is in agreement with the fact that the observed maximum Doppler velocity of the peak of the line profile is \mbox{$400 \lesssim \vert V \vert\lesssim 450$~km\,s$^{-1}$} (see \Cref{fig:vel}b), which is comparable to the primary wind terminal velocity of 420~km\,s$^{-1}$ \citep{2012MNRAS.423.1623G}. This region is also favored by arguments related to the energy required to produce the observed line luminosity, as extensively discussed in previous works \citep[e.g.][]{2006ApJ...640..474M,2011ApJ...740...80M,2012ApJ...746...73T}. Thus, in general, any feasible scenario for the \heii~production around periastron passage requires that the emitting region is adjacent to the wind-wind collision shock cone, close to its apex. This is the basic assumption for the model we propose next.

\subsection{A model for the variations in the {\rm He}\,{\sc ii}~$\lambda4686$ \btxt{equivalent width curve}: opacity and geometry effects.}

\btxt{Assuming that the He\,{\sc ii}~$\lambda4686$ emission is produced close to the apex of the shock cone, variations in the observed emission can, in principle, be explained by the increase in the total opacity along the line of sight to the emitting region, as the secondary moves deeper inside the dense primary wind}. Intuitively, this mechanism would cause a gradual decrease (or increase, after periastron passage) of the observed flux that would depend on the extent and physical properties of the optically thick region in the extended primary wind, and also on the orbital orientation to the observer.

\begin{figure}
  \centering
  \includegraphics[width=\linewidth]{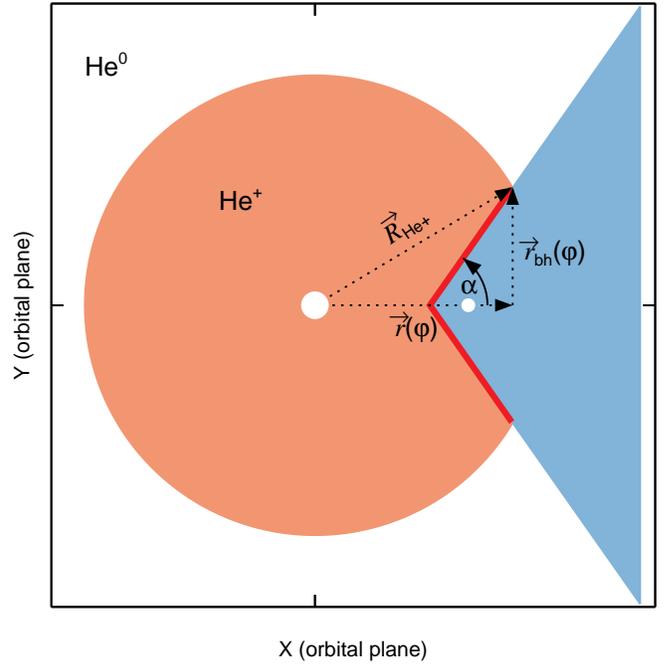}
  \caption{Sketch (not to scale) of the intersection between the wind-wind interacting region (represented by a cone) and the He$^+$ region (represented by the orange circle) at phases around periastron. When the stars (white circles) are close enough, so that the wind-wind interacting region penetrates the He$^+$ core in the primary wind, high-energy radiation produced in the shock cone (red line) can easily ionize He$^+$ ions, and some of the radiation due to the eventual recombination processes will escape through the aperture of the `bore hole'. This will lead to an increasing in the production of \heii~photons.\label{fig:bhgeometry}}
\end{figure}

This same approach was used by \citet{2008MNRAS.388L..39O} to show that the overall behavior of the RXTE X-ray light curve can be reproduced by assuming that the X-ray emission comes from a point source located at the apex of the shock cone and prone to attenuation by the primary wind. Recently, \citet{2014ApJ...784..125H} suggested that the variations across the spectroscopic events are composed of a combination of \mbox{(1) occultation} of the X-ray emitting region by the extended primary wind and \mbox{(2) decline} of the X-ray emissivity at the apex. In any case, opacity effects (either attenuation or occultation) can play an important role on the observed intensity of the radiation. Since the X-ray light curve shares some similarities with the He\,{\sc ii}~$\lambda4686$ \btxt{equivalent width curve} (both rise to a maximum before falling to a minimum when there is no emission at all), we tested the hypothesis that the variations in the \heii~equivalent width could also be the result of intrinsic emission attenuated by the extended primary wind.

\begin{figure}
  \centering
  \includegraphics[width=\linewidth]{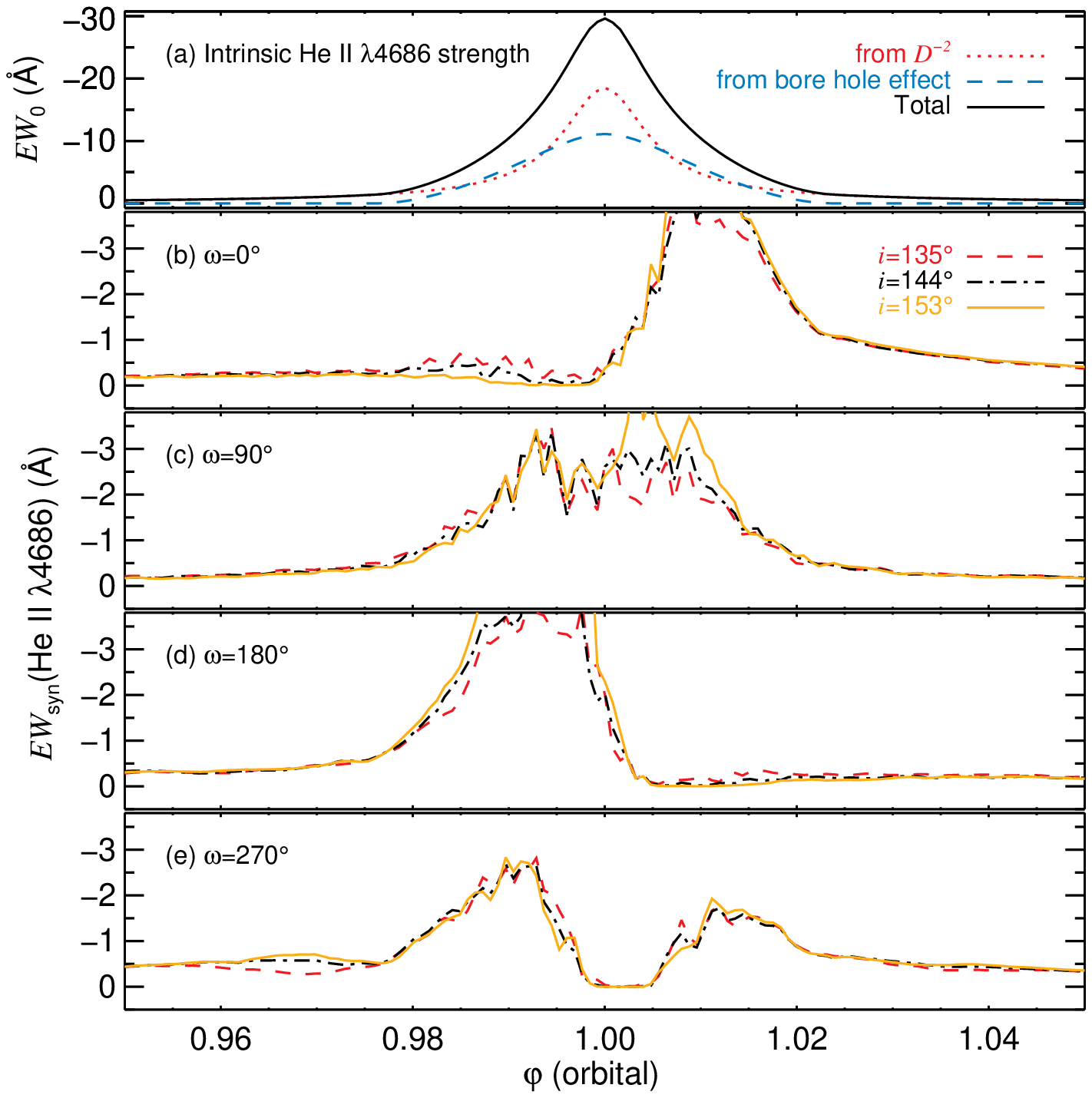}
  \caption{\btxt{(a) Intrinsic and (b--e) synthetic He\,{\sc ii}~$\lambda4686$ equivalent width curves}. The basic assumption is that the He\,{\sc ii}~$\lambda4686$ emission comes from a region located at the apex and varies accordingly to the total column density in the line of sight to it as derived from 3D SPH simulations \btxt{($D$ is the distance between the stars). The same intrinsic line strength was assumed for all orbit orientations.} For each value of the longitude of periastron ($\omega$), the different lines show the expected behavior of the equivalent width for the orbit inclinations ($i$) indicated in the legend. \btxt{Variations in the equivalent width curve} are more sensitive to $\omega$ than $i$. Based on these plots and the observations, we can promptly exclude $0\degr\lesssim\omega\lesssim180\degr$ and favor $\omega$ close to $270\degr$.\label{fig:models}}
\end{figure}

\btxt{We used \mbox{3D SPH} simulations of $\eta$~Car from \citet{2013MNRAS.436.3820M} to calculate the total optical depth in the line of sight to the apex at each phase by using}
\begin{equation}
		\tau_{\rm A}(\varphi) = \kappa_e \, \int_{z_0(\varphi)}^{R} \rho \, dz,
	\label{eq:tau}
\end{equation}
where $\rho$ and $\kappa_e$ are, respectively, the mass density and the mass absorption coefficient of the material in the line of sight to the apex. The integration starts at the position of the apex at each phase $z_0(\varphi)$ and goes up to the boundaries of the 3D SPH simulations, which, in this case, is a sphere with radius \mbox{$R=154.5$~AU}. For the present work, we assumed that electron scattering is the dominant process for the attenuation of the radiation in the line of sight, which corresponds to \mbox{$\kappa_e=0.34$~cm$^2$\,g$^{-1}$}. Therefore, under these circumstances, the synthetic equivalent width, $EW_{\rm syn}$, at each orbital phase, was obtained using
\begin{equation}
	EW_{\rm syn}(\varphi) = EW_0(\varphi) \, e^{-\tau_{\rm A}(\varphi)},
	\label{eq:ew}
\end{equation}
where $EW_0(\varphi)$ is the intrinsic equivalent width (corresponding to the unattenuated flux). In the present work, we included only two mechanisms responsible for $EW_0$: (i) \btxt{a continuous production of \heii~photons that varies reciprocally with the square of the distance $D$ between the stars} \citep[\btxt{i.e. in radiative conditions}; see][]{Fahed:2011gh} and (ii) an additional, temporary contribution from the `bore hole' effect that depends on how deep the apex of the shock cone penetrates inside the primary wind \citep[see][]{2010RMxAC..38...52M}.

We included the contribution from the `bore hole' effect because, near periastron, due to the highly eccentric orbit, the wind-wind interacting region penetrates into the inner regions of the primary wind, eventually exposing its He$^+$ core\footnote{The He$^+$ core has a radius of about 3~AU \citep[][]{2001ApJ...553..837H,2012MNRAS.423.1623G}. Assuming an eccentricity of 0.9, the apex should be inside the He$^+$ core for $0.98 \lesssim \varphi({\rm orbital}) \lesssim 1.02$.}. The contribution from this mechanism to the observed \heii~flux is proportional to how large the `bore hole' is (see \Cref{fig:bhgeometry}). This assumption relies on the fact that high-energy radiation produced in the shock cone inside the He$^+$ region (the red line between the cone and the sphere in \Cref{fig:bhgeometry}) can create He$^{++}$ ions, whose recombination will produce \heii~photons. Some will eventually escape through the `bore hole' and be detected by the observer.

The radius $r_{\rm bh}$ of the `bore hole' is wavelength-dependent and varies with orbital phase. Considering the He$^+$ region, $r_{\rm bh}$ as a function of the orbital phase is given by
\begin{equation}
	r_{\rm bh}(\varphi) = (R^2_{\rm He^+} - r^2(\varphi))^{1/2},
	\label{eq:bh1}
\end{equation}
where $r(\varphi)$ is the distance between the primary star and the plane formed by the aperture of the `bore hole' (see \Cref{fig:bhgeometry}), given by
\begin{equation}
	r(\varphi) = \frac{c^2\,z_0(\varphi)+(c^2\,(R^2_{\rm He^+} - z^2_0(\varphi)) + R^2_{\rm He^+})^{1/2}}{c^2 + 1},
	\label{eq:bh2}
\end{equation}
and $c=\tan\alpha$, where $\alpha$ is the half-opening angle of the cone formed by the wind-wind interacting region. In \Cref{eq:bh1,eq:bh2}, $R_{\rm He^+}$ is the radius of the He$^+$ region in the primary wind and $z_0(\varphi)$ is the distance between the primary and the apex at a given orbital phase. \btxt{\Cref{fig:models}a shows the contribution from each mechanism to the total intrinsic equivalent width $EW_0$. The relative contribution was chosen so that the transition between the two regimes occurred smoothly, as required by the observations. Thus, by combining the intrinsic strength for the line emission with the total opacity in the line of sight, we were able to calculate a synthetic equivalent width curve for different orbit orientations. The results for selected orbital orientations are shown in \mbox{\Cref{fig:models}b--e}.}

\begin{figure}
  \centering
  \includegraphics[width=\linewidth]{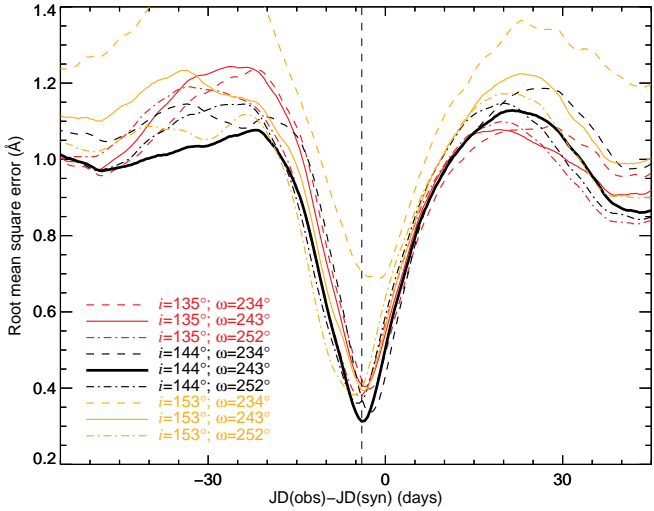}
  \caption{Examples of the result of the comparison between the observed He\,{\sc ii}~$\lambda4686$ \btxt{equivalent width curve and a series of synthetic curves} from 3D SPH simulations with different orbital orientations (3 inclinations and 3 longitude of periastron). \btxt{For each comparison, we shifted the models in time to determine the time of periastron passage. Each curve shows the least root mean square error as a function of the difference between the time of the observations, JD(obs), and that of the synthetic equivalent width curve, JD(syn), for the indicated orbit orientation}. The best match occurred for $i=144\degr$, $\omega=243\degr$ (black solid line), and a time shift of about $-4.0$~d.\label{fig:residuals}}
\end{figure}

\subsection{Modeling the {\rm He}\,{\sc ii}~$\lambda4686$ equivalent width}

\subsubsection{The direct view of the central source}

Based on the comparison between the overall profile of the observed \heii~equivalent width curve from the past 3 cycles and those synthetic curves shown in \Cref{fig:models}, one can readily discard models with $0\degr \lesssim \omega \lesssim 180\degr$. Orientations with $\omega=0\degr$ produce results that have excessively high optical depths before periastron passage and way too little after it. This orientation cannot reproduce the observed rise of the equivalent width before periastron passage and also overestimates its strength after it. Orientations with $\omega=90\degr$ produce symmetrical profiles, which do not correspond to the observations. Orientations with $\omega=180\degr$ can reproduce fairly well the observations before periastron but fail to reproduce the observed equivalent width after periastron (they underestimate P3).

Regardless of the overall profile of the synthetic equivalent width curves, the crucial problem of models with $0\degr \lesssim \omega \lesssim 180\degr$ is that they cannot reproduce the observed phase of the deep minimum -- the week-long phase where the observed equivalent width is zero.

Orientations with $230\degr \lesssim \omega \lesssim 270\degr$, on the other hand, seem to provide a good overall profile, as they predict an increase just before periastron passage followed by a rapid decrease to zero right after periastron passage, and the return to a lower (in modulo) equivalent width peak before fading away (P3). Thus, we focused our analysis on $\omega$ in this range.

The duration of the interval when the synthetic equivalent width remains near zero (and whether it is ever reached) is also regulated by the orbital inclination. Thus, we compared the observations with 16 synthetic \btxt{equivalent width curves} obtained from the permutation of 4 values of orbital inclination (\mbox{$i\in\{126\degr,135\degr,144\degr,153\degr\}$}) and 5 values of longitude of periastron \mbox{($\omega\in\{225\degr,234\degr,243\degr,252\degr,261\degr\}$}). Then we calculated the root mean square error (RMSE) between each model and the observations. The values for $i$ and $\omega$ were obtained from a pre-defined grid within the 3D SPH models. Also, note that the orbital plane is parallel to the plane of the sky for $i=0\degr$ or $i=180\degr$, whereas for $i=90\degr$ or $i=270\degr$ they are perpendicular to each other. Thus, the set of orbital inclinations that we investigated in this work was chosen based on the premise that the orbital axis is aligned with the Homunculus polar axis \cite[see][]{2012MNRAS.420.2064M}.

\begin{figure}
  \centering
  \includegraphics[width=\linewidth]{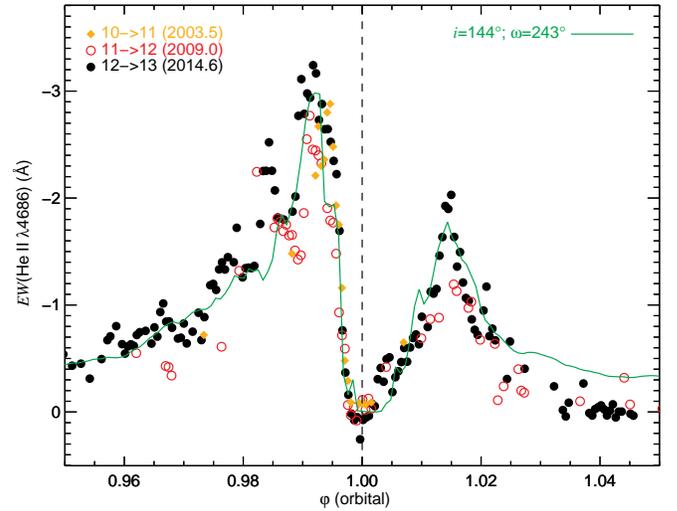}
  \caption{\btxt{Folded He\,{\sc ii}~$\lambda4686$ equivalent width curve compared with the best synthetic model (solid green line). The minimum RMSE occurs when the synthetic model is shifted by $-4.0$~d relative to the observations. Note that now the phase corresponds to the orbital phase}.\label{fig:folded}}
\end{figure}

For each model, we also searched for the time shift to be applied to the \btxt{models} that would result in the least root mean square value between the model and the observations. Examples of the results of this analysis are shown in \Cref{fig:residuals}. \btxt{The minimum RMSE was reached for an orbit orientation with \{$i=144\degr$, $\omega=243\degr$\} and a time shift \mbox{$\mathrm{JD(obs)}-\mathrm{JD(syn)} = -3.5$~d}}. A comparison between the best model \btxt{(with the derived time shift applied)} and the observations is shown in \Cref{fig:folded}.

Regarding the mean value and uncertainty of these results, statistical analysis showed that there are no significant differences between a model with a combination of $i=144\degr$ and $234\degr \lesssim \omega \lesssim 252\degr$. In fact, within the range of orbit orientations that we focused our analysis on, only these models resulted in RMSE significantly lower than the others at the $2\sigma$ level. Therefore, the mean values we adopted for $i$ and $\omega$ are, respectively, $144\degr$ and $243\degr$ (coincidently equal to the best match), whereas the uncertainty on both values is defined by the step in the number of lines of sight used to produce the synthetic \btxt{equivalent width curves} from the 3D SPH models, which was set to $9\degr$. In order to estimate the mean and uncertainty on the time shift, we adopted a sample composed of the time shift that resulted in the least RMSE for each one of the 16 models. The result was a mean time shift of $-4.0\pm2.0$~d (the standard error is $\pm0.5$~d). This means that periastron passage occurs 4~d after the onset of the \heii~deep minimum.

\subsubsection{The event as viewed from high stellar latitudes}

An independent way to verify the reliability of our results would be analyzing the \heii~\btxt{equivalent width curve} from different viewing angles and comparing them with the results from the direct view of the central source. Fortunately, in the case of $\eta$~Car, this is possible due to the bipolar reflection nebula -- the Homunculus nebula -- that surrounds the binary system. Each position along the Homunculus nebula `sees' the central source along a different viewing angle \citep{2003ApJ...586..432S}. An interesting position is the FOS4 \citep[e.g.][]{1995AJ....109.1784D,1999ASPC..179..107H,1999A&A...344..211Z,2001ASPC..242...29R,2005A&A...435..303S}, a region about \mbox{$1$~arcsec$^2$} in area located approximately 4.5~arcsec from the central source along the major axis of the Homunculus\footnote{As remarked by \citet{2005A&A...435..303S}, the initial definition of FOS4, done using \textit{HST} Faint Object Spectrograph images obtained in 1996-97, was a $0.5$~arcsec wide region located $4.03$~arcsec from the central source at \mbox{${\rm PA}=135\degr$}. Given that the Homunculus nebula expands at an average rate of $0.03$~arcsec\,yr$^{-1}$ \citep[e.g.][]{1998AJ....116..823S}, the current distance between FOS4 and the central source is about $4.5$~arcsec} that is believed to reflect the spectrum originating at stellar latitudes close to the polar region \citep[for comparison, the spectrum from the direct view of the central source seems to arise from intermediate stellar latitudes; see e.g.][]{2003ApJ...586..432S}.

\begin{figure}
  \centering
  \includegraphics[width=\linewidth]{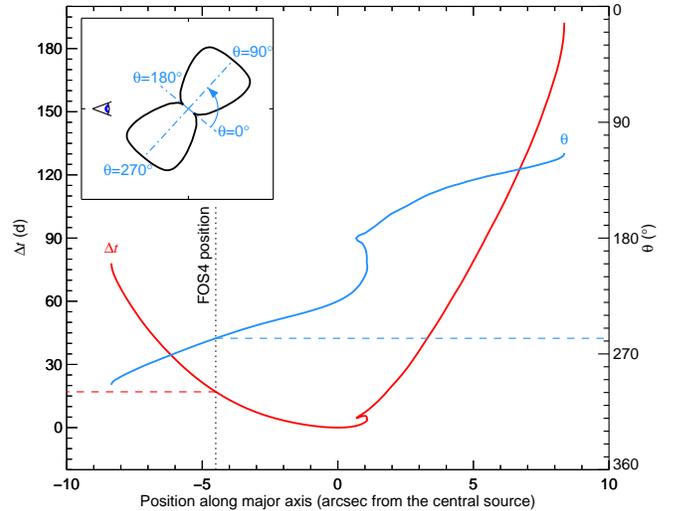}
  \caption{Time delay ($\Delta t$) and stellar latitude ($\theta$) as a function of the projected position along the major axis of the Homunculus nebula. The inset illustration shows the the definition of the stellar latitude regarding the orientation of the Homunculus nebula. Values between $0\degr$ and $180\degr$ correspond to the NW lobe ($\theta=90\degr$ being the NW pole), whereas latitudes between $180\degr$ and $360\degr$ correspond to the SE lobe ($\theta=270\degr$ is the SE pole). In this system, the direct view of the central source corresponds to \mbox{$\theta \equiv \theta_0=229\degr$}. The vertical dotted line indicates the FOS4 position, whereas the horizontal dashed lines indicate the corresponding values for $\Delta t=17$~d (left vertical axis) and $\theta=258\degr$ (right vertical axis).\label{fig:hom_model}}
\end{figure}

We compared our results with those obtained at FOS4 by \citet{2015arXiv150404940M} using the same analysis that we did for the direct view of the central source. Note, however, that the time shift obtained from the observations at FOS4 needs to be corrected by the light travel time and the expansion of the reflecting nebula. We achieved this by using the Homunculus model derived by \citet{2006ApJ...644.1151S}, to determine the necessary parameters of FOS4 (see \Cref{fig:hom_model}). In a coordinate system where a stellar latitude of $\theta=90\degr$ corresponds to the pole of the receding NW lobe and $\theta=270\degr$ corresponds to the pole of the approaching SE lobe, the spectrum reflected at FOS4 corresponds to a stellar latitude of $\theta({\rm FOS4})=258\degr$, whereas the central source is viewed at $\theta_0=229\degr$.

\begin{figure}
  \centering
  \includegraphics[width=\linewidth]{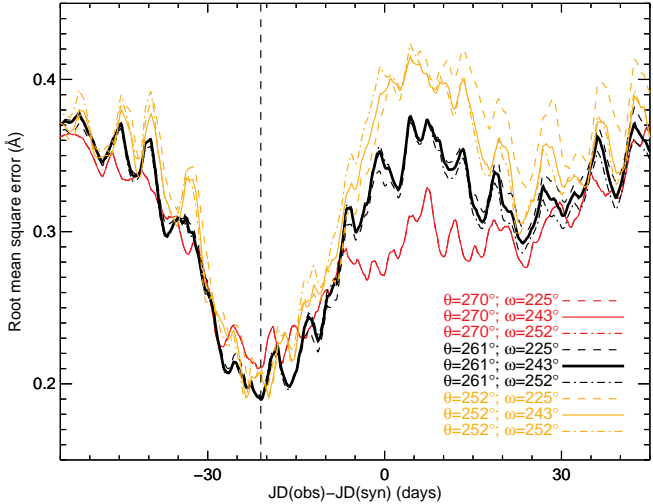}
  \caption{Same as \Cref{fig:residuals}, but for the position FOS4. At this position, we are looking at a spectrum originated at a stellar latitude $\theta$ that is being reflected off the Homunculus nebula. In this system, $\theta=90\degr$ corresponds to the NW lobe pole, whereas \mbox{$\theta=270\degr$} corresponds to the SE lobe pole. The best match (black solid line) occurred for \mbox{$\theta=261\degr$}, \mbox{$\omega=243\degr$}, and a time shift of $-21.0$~d.\label{fig:residuals2}}
\end{figure}

\Cref{fig:residuals2} shows the search for the best match between the models and the observations. Since the orbital axis is closely aligned with the Homunculus polar axis, we investigated models corresponding to stellar latitudes $\theta\in \{252\degr, 261\degr, 270\degr\}$ and $\omega\in\{225\degr, 243\degr, 252\degr\}$. The best match was found for \{$\theta=261\degr$, $\omega=243\degr$\} and a time shift of $-21.0$~d. This model resulted in a RMSE significantly lower (at the $2\sigma$ level) than any other one. \btxt{Note that this time shift of 21~d, obtained empirically \cite[][]{2006ApJ...644.1151S}, is consistent with previous estimates by \cite{2005A&A...435..303S} and \cite{2011ApJ...740...80M} based on geometrical arguments}.

\btxt{Since the observations at FOS4 were not as frequent as for the direct view, our analysis is subject to aliasing}, which explains the high frequency oscillations observed in \Cref{fig:residuals2}. Despite that, we determined a mean time shift at FOS4 of $-21.6\pm1.3$~d (the standard error is $0.4$~d). Note that the smaller uncertainty of this result, when compared with the direct view of the central source, is not real. This is just an effect introduced by the fact that we used a smaller sample for the trial models for the polar region than for the direct view. Also, the variations between the trial models for the polar region are not as large as the ones for the direct view, which reduces the dispersion of the minimum RMSE (all the models for the pole region have similar time shift).

\begin{figure}
  \centering
  \includegraphics[width=\linewidth]{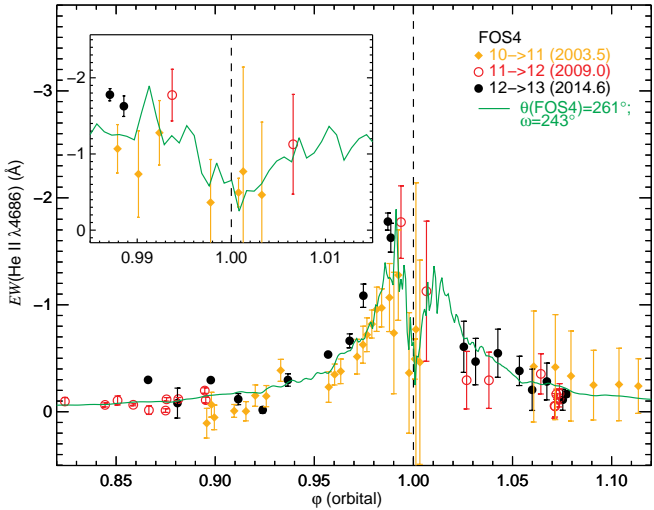}
  \caption{FOS4 observations from \protect \citet{2015arXiv150404940M} corrected by the time delay ($17$~d). The green solid line corresponds to the best match model: $\theta({\rm FOS4})=261\degr$ and $\omega=243\degr$. The vertical dashed line indicates periastron passage.\label{fig:folded2}}
\end{figure}

The time shift we derived for FOS4 has to be corrected by light travel delay. Considering that the lobes are expanding at $650$~km\,s$^{-1}$, the spectrum reflected at the FOS4 position would be delayed by $\Delta t=17$~d relative to the direct view of the central source. This means that $17$~d, out of the observed $-21.6$~d, are due to light travel effect that must be taken into account in order to compare with the observations of the central source (again, the minus sign means periastron occurs after the onset of the \heii~deep minimum). Therefore, the best model for the observations at FOS4 results in an effective time shift of \mbox{$17$~d$-21.6$~d$=-4.6\pm0.4$~d}, which is in agreement with the results obtained from the direct view of the central source. \Cref{fig:folded2} shows the best model compared with the observations at FOS4 (corrected for light travel time). An interesting result is that when the `bore hole' effect is included, the models overestimate the equivalent width at FOS4, which suggests that the contributions due to this mechanism is small at high stellar latitudes. This makes sense, given the fact that the contribution from the `bore hole' effect will be significant toward where the cavity is pointing to (i.e. low and intermediate latitudes).

The net result is that the ephemeris equation derived from the reflected observations at FOS4 (high stellar latitude), after correction by the travel time delay, is the same as for the direct view (intermediate stellar latitude). Therefore, the variations across the event for the \heii~seem to be ultimately determined by the high opacity in the line of sight to the \heii~emitting region during periastron passage, and not by a decrease in the intrinsic emission.

\subsection{Ephemeris equation for periastron passage}
\begin{figure}
  \centering
  \includegraphics[width=\linewidth]{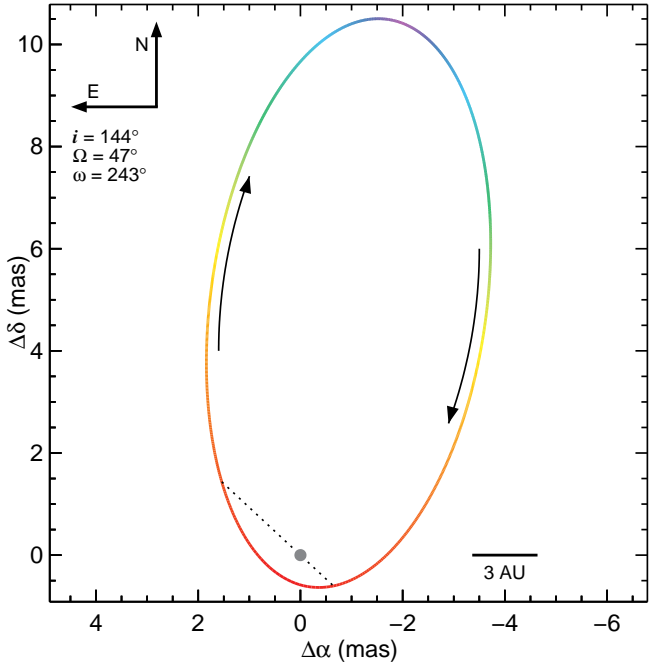}
  \caption{Orientation of the orbit of the secondary star as projected onto the sky using the mean value of the orbital elements derived in the present work. The arrows indicate the direction of the orbital motion. Note that the orbit is with respect to the primary star. The orbit was color-coded using the actual (not projected) distance between the stars, so that the orbit path becomes red as the stars are close to each other around periastron passage, and purple when they are at apastron. We adopted a distance of $2350$~pc \cite[][]{2006ApJ...644.1151S}.\label{fig:orbit}}
\end{figure}

The results presented in this paper allowed us to determine the time of the periastron passage, $T_0(2014.6)$, from two different line of sights. For the direct view, \mbox{$T_0(2014.6)=2456874.1$ (standard deviation: $\pm2.8$~d)}, whereas for the FOS4 \mbox{$T_0(2014.6)=2456874.7$} (standard deviation: $\pm2.3$~d). Therefore, the time of periastron passage is given by
\begin{equation}
	{\rm JD (periastron~passage)} = 2456874.4 + 2022.7\,E,
	\label{eq:eph}
\end{equation}
where $E=(\Phi - 13)$, $\Phi$ is the mean anomaly plus cycle\footnote{Following \cite{2004IBVS.5492....1G}, the cycle counting starts on \mbox{${\rm JD}=2430578.4$}, corresponding to \mbox{1942 Aug 06}, around the time when the first observed `low-excitation state' due to a spectroscopic event was reported by \citet{1953ApJ...118..234G}.}, and $2022.7$ corresponds to the mean period of the spectroscopic cycle. The time of periastron passage determined by \Cref{eq:eph} has an uncertainty of $\pm2.0$~d (standard error), which is the result of error propagation from all the parameters used to determine the mean values shown in that equation. \Cref{tab:orbitelements} summarizes all the orbital elements derived in this work and \Cref{fig:orbit} illustrates the orientation of the orbit projected onto the sky using the mean value for the orbital elements. 

\subsection{The {\rm He}\,{\sc ii}~$\lambda4686$/X-ray connection}
The comparison between the \heii~emission and the X-rays is inevitable because both are tracers of high-energy processes that might be interconnected. For example, X-rays produced in the wind-wind shock might be used to doubly-ionize neutral helium, in which case the \heii~emission should vary in a similar way to the X-rays.

In $\eta$~Car, the onset of the deep minimum, as determined from the observed \heii~and $2$--$10$~keV X-ray emission light curves, occurs almost at the same time \cite[see e.g.][]{2012ApJ...746...73T,2014ApJ...784..125H}. Incidentally, no significant changes in the time of the start of the deep minimum were observed over the past 3 cycles (\bbtxt{Corcoran et al. 2016}, in preparation). However, the recovery of the \heii~emission and the $2$--$10$~keV X-ray flux occurs in a completely different way. So far, the observations support a scenario where the \heii~emission presents a recovery phase (after the deep minimum) that is stable and repeatable, whereas for the X-rays, recovery cannot be predicted. Indeed, in 1998.0 and 2003.5, the full recovery of the $2$--$10$~keV X-ray flux occurred about 90~d after the onset of the deep minimum, but in 2009.0 it happened 30~d earlier than that. Even more intriguing, in 2014.6, the observations indicate that the recovery was intermediate between the 1998.0/2003.5 and 2009.0. Thus, the question is why does the \heii~and the X-ray emission enter the deep minimum at the same time but recover from it so differently? Are they connected?

\cite{2014ApJ...784..125H} showed that the $2$--$10$~keV X-ray flux can be split into two regimes that behave differently in time: the soft (2--4~keV) and the hard X-ray band (4--8~keV). Those authors showed that the hard X-ray flux (4--8~keV) drops to a minimum before the soft X-ray flux (2--4~keV). In fact, by the time the latter reaches the minimum, the former is already recovering from it. Interestingly, \heii~emission enters the minimum phase about 4~d \textit{after} the hard X-rays and about 12~d \textit{before} the soft X-rays. This fact not only corroborates the idea that the \heii~emitting region is located in the vicinity of the apex (likely close to the hard X-ray emitting region), but is also consistent with the scenario proposed in this paper, in which the onset of the deep minimum is regulated by the orbital orientation and opacity effects, as the secondary moves toward periastron. In this scenario, emission from the apex should disappear first, followed by emission produced in regions gradually farther away from the WWC apex (i.e., downstream of the WWC structure). Therefore, the time when the deep minimum starts should be approximately the same for all features arising from spatially adjacent regions, as might be the case for \heii~and the X-rays (especially the hard band flux).

\begin{table}

    \centering
    \caption{Orbital elements derived in the present work, assuming orbital eccentricity $\epsilon=0.9$.}\label{tab:orbitelements}
    \begin{tabular}{lcr}

    \hline\hline
    Parameter & Symbol & Value or range \\\hline
     Inclination & $i$ & $135\degr$--$153\degr$ \\
     Longitude of periastron & $\omega$ & $234\degr$--$252\degr$ \\
     Period & $P$ & $2022.7\pm0.3$\textsuperscript{a} \\
     Time of periastron passage & $T_0$(2014.6) & $2456874.4\pm1.3$\textsuperscript{a} \\\hline
     \multicolumn{3}{l}{\textsuperscript{a}\footnotesize{The error is the standard error of the mean.}}\\
    \end{tabular}

\end{table}

Using the ephemeris from \cite{2014ApJ...784..125H}, the deep minimum phase in the hard X-rays occurred, approximately, from \mbox{$\mathrm{JD}=2454843$ through $\mathrm{JD}=2454859$}, which suggests that superior conjunction would have occurred within this interval. \btxt{According} to \Cref{eq:eph}, the 2009.0 periastron passage occurred on $\mathrm{JD}=2454851.7$, which falls approximately in the middle of the phase of minimum hard X-ray flux. Moreover, note that, for $\omega=243\degr$, superior conjunction occurs very close to periastron passage, only 3.6~d later. Therefore, the orbital orientation as derived from X-ray and \heii~emission \btxt{are consistent}. This is not surprising in the scenario we propose here. Indeed, close to periastron passages, we expect the hard X-ray and the \heii~to behave similarly (although they might not have a causal relation) because the apex of the shock cone is inside the primary He$^+$ core where the \heii~emission is being produced at these phases. The soft X-rays, however, are produced in extended regions along the shock cone, and will eventually be blocked by the inner dense parts of the primary star at a later time.

After periastron passage, the \heii~emission behaves differently from the X-rays likely due to intrinsic physical processes that inhibit or even shut off the latter, but not the former. For example, if the shock cone structure would switch from adiabatic to radiative during periastron passage, the result would be a substantial decrease of the hardest X-ray emission because the wind-wind interacting region would be cooler. The subsequent recovery of the X-ray emission would then depend on stochastic processes in order to re-establish the emission from the hot plasma in the wind-wind interaction region. This is not, however, the case for the \heii~emission because, during periastron passage, the He$^+$ region of the primary star is exposed by the wind-wind interacting region, and, therefore, \heii~emission can still be produced during this phase, even if there is very low X-ray emission from the WWC region.

\section{Final remarks}\label{sec:remarks}
\btxt{Analysis of the data we have thus far indicate that only P2 has significantly increased in strength over the past two cycles (2009.0 and 2014.6). This is a robust result because we have daily measurements during the appearance of P2 for those epochs. However, except during the 2014.6 periastron passage, P1 and P3 have never been monitored with high cadence, which makes the comparison rather difficult because the equivalent width of the He\,{\sc ii}~$\lambda4686$ line shows variations at a wide range of timescales.}

\btxt{One clear example is the absolute maximum strength of P3. The results from the 2009.0 analysis (including the additional data obtained with Hexapod/BESO) suggest that it is composed of a broad peak with a maximum absolute value of about $1.3$~\AA, but the 2014.6 results indicate that P3 is actually a combination of broad and sharp components, with a maximum absolute value of $2$~\AA. The broad component does repeat from cycle to cycle, but we cannot conclude the same for the narrow component, as we did not have enough time coverage during the time the narrow component seems to appear. Nonetheless, a caveat here is that short-timescale variations (less than a week) might occur due to local stochastic mechanisms, like the flares seen  in X-rays before periastron passages. Such fluctuations cannot be easily distinguished from cyclic variations.}

\btxt{Our model is especially sensitive to two parameters: (\textit{i}) the total opacity in the line of sight to the apex and (\textit{ii}) to the size of the He$^+$ region in the primary wind. Hence, changes in these parameters will reflect on the overall behavior of the He\,{\sc ii}~$\lambda4686$ equivalent width curve. As a matter of fact, both parameters are ultimately connected to the mass-loss rate of the primary star. Changes in the primary's mass-loss rate would result in variations in the timing and strength of the \heii~\btxt{equivalent width curve} \cite[as already discussed in][]{2013MNRAS.436.3820M}. The repeatability of the overall behavior of the \heii~equivalent width over the past three cycles, as shown in this work, corroborates previous results that rule out large changes in mass-loss rate from the primary star over that time interval.}

\section{Conclusions}\label{sec:conclusions}

We have monitored the \heii~emission line across the 2014.6 event using many ground-based telescopes as well as the \textit{HST}. The main results derived from the analysis of the collected data are listed below.

\begin{itemize}
\item The period of $2022.9\pm1.6$~d, derived from He\,{\sc ii}~$\lambda4686$ monitoring, is in agreement with previous results;

\item Based on several different measurements across the electromagnetic spectrum, the mean orbital period is $2022.7\pm0.3$~d;

\item We have not detected statistically significant changes in the overall behavior of the He\,{\sc ii}~$\lambda4686$ \btxt{equivalent width curve} when comparing the events of 2009.0 and 2014.6 (best time sampled), which implies that the mechanism behind the production of EUV/soft X-rays photons must be relatively stable and recurrent;

\item When comparing each peak separately, between 2009.0 and 2014.6, P1 and \bbtxt{the broad component of} P3 have not changed significantly. Nevertheless, \btxt{P2 has increased by 26\%};

\item We have proposed a model to explain the variations of the \heii~equivalent width curve across each event. The model assumes two different mechanisms responsible for the intrinsic production of the line emission: (1) a component that \btxt{is inversely proportional to the square of the distance between the two stars} (always present throughout the orbit) and (2) another one associated with the `bore hole' effect (present within about $30$~d before and $30$~d after periastron passage; negligible outside this time interval). The intrinsic emission was then convolved with the total optical depth in the line of sight to the \heii~emitting region (computed from 3D SPH simulations) to create a synthetic equivalent width that was compared with the observations;

\item Our model was able to successfully reproduce the overall behavior of the \heii~\btxt{equivalent width curve} from two very different viewing angles: a direct view and a polar view of the central source. The latter was possible due to observations from the FOS4 position on the SE lobe of the Homunculus nebula. This particular position is known for its capability of reflecting the spectrum of the central source produced at high stellar latitudes;

\item The best match between the models and observations from two different viewing angles (direct view of the central source and FOS4) suggests \mbox{$234\degr \lesssim \omega \lesssim 252\degr$};

\item We have determined the time of periastron passage by comparing the time shift required to give the best match between models (for which we know the orbital phase) and observations for two different lines of sight. The results suggest that both directions `see' the periastron passage at the same time, about $4$~d after the start of the deep minimum (as seen in the direct view of the central source);

\end{itemize}

In summary, our results suggest that the variations observed in the \heii~equivalent width curve across the spectroscopic events are governed by a combination of the orbit orientation regarding the observer and the total optical depth in the line of sight to the emitting region. This is not a classical eclipse of the emitting region by the primary wind (the so-called `wind eclipse' scenario). Instead, the ultimate nature of the spectroscopic event can be ascribed to the deep burial of the secondary star in the densest parts of the primary wind, so that emission from the vicinity of the apex of the shock cone cannot easily escape the system, resulting in a temporary decrease of the observed emission regardless of line of sight.

As a reminder, the onset of the next \heii~deep minimum will occur on \mbox{2020 February 13}. Periastron passage should occur four days later.

\section*{Acknowledgments}
During part ot this research, M.~T. was supported by CNPq/MCT-Brazil through grant 201978/2012-1. A.~D. thanks FAPESP for financial support through grant 2011/51680-6. F.~W. thanks K.~Davidson and R.~Humphreys for their discussions leading to part of these observations. \btxt{Some of the spectra were obtained under the aegis of Stony Brook University, whose participation has been supported by the office of the Provost.} We are grateful to Pam~Kilmartin and Fraser~Gunn for helping with the observations at MJUO. A.~F.~J.~M. is grateful to financial aid to NSERC (Canada) and FRQNT (Qu\'ebec). N.~D.~R. gratefully acknowledges his CRAQ (Qu\'ebec) postdoctoral fellowship. \btxt{D.~J.~H.} acknowledges support from \mbox{HST-GO 12508.02-A}. T.~I.~M. and C.~M.~P.~R. are supported by an appointment to the NASA Postdoctoral Program at the Goddard Space Flight Center, administered by Oak Ridge Associated Universities through a contract with NASA. We are grateful to STScI for support with the observational schedule. This publication is based (in part) on spectroscopic data obtained through the collaborative Southern Astro Spectroscopy Email Ring (SASER) group.This research has made extensive use of NASA's Astrophysics Data System, \href{http://idlastro.gsfc.nasa.gov}{{\sc idl} Astronomy User's Library}, and David Fanning's \href{http://www.idlcoyote.com/index.html}{{\sc idl} Coyote library}. This research has made use of the \href{http://exoplanetarchive.ipac.caltech.edu/index.html}{NASA Exoplanet Archive}, which is operated by the California Institute of Technology, under contract with the National Aeronautics and Space Administration under the Exoplanet Exploration Program. \btxt{We would like to thank an anonymous referee for constructive suggestions that improved the presentation of this work}.

{\it Facilities:} \facility{CTIO:1.5m (CHIRON)}, \facility{LNA:ZJ0.6m (Lhires {\sc iii})}, \facility{LNA:1.6m (Coud\'e)}, \facility{SOAR (Goodman)}, \facility{MtJohn:1m (HERCULES)}, \facility{CASLEO (REOSC DC), \facility{\bbtxt{OCA:Hexapod (BESO)}}}

\bibliographystyle{mn2e}

\begin{thebibliography}{}
\makeatletter
\relax
\def\mn@urlcharsother{\let\do\@makeother \do\$\do\&\do\#\do\^\do\_\do\%\do\~}
\def\mn@doi{\begingroup\mn@urlcharsother \@ifnextchar[{\mn@doi@}{\mn@doi@[]}}
\def\mn@doi@[#1]#2{\def\@tempa{#1}\ifx\@tempa\@empty
  \href{http://dx.doi.org/#2}{doi:#2}\else \href{http://dx.doi.org/#2}{#1}\fi
  \endgroup}
\def\mn@eprint#1#2{\mn@eprint@#1:#2::\@nil}
\def\mn@eprint@arXiv#1{\href{http://arxiv.org/abs/#1}{{\tt arXiv:#1}}}
\def\mn@eprint@dblp#1{\href{http://dblp.uni-trier.de/rec/bibtex/#1.xml}{dblp:#1}}
\def\mn@eprint@#1:#2:#3:#4\@nil{\def\@tempa {#1}\def\@tempb {#2}\def\@tempc
  {#3}\ifx \@tempc \@empty \let\@tempc\@tempb \let\@tempb\@tempa \fi \ifx
  \@tempb \@empty \def\@tempb{arXiv}\fi \@ifundefined
  {mn@eprint@\@tempb}{\@tempb:\@tempc}{\expandafter \expandafter \csname
  mn@eprint@\@tempb\endcsname \expandafter{\@tempc}}}

\bibitem[\protect\citeauthoryear{Abraham \& Falceta-Gon{\c c}alves}{Abraham \&
  Falceta-Gon{\c c}alves}{2007}]{2007MNRAS.378..309A}
Abraham Z.,  Falceta-Gon{\c c}alves D.,  2007, Monthly Notices of the Royal
  Astronomical Society, 378, 309

\bibitem[\protect\citeauthoryear{Bauer}{Bauer}{2012}]{Bauer:2012ef}
Bauer D.~F.,  2012, Journal of the American Statistical Association, 67, 687

\bibitem[\protect\citeauthoryear{Birnbaum \& Tingey}{Birnbaum \&
  Tingey}{1951}]{Birnbaum:1951fy}
Birnbaum Z.~W.,  Tingey F.~H.,  1951, The Annals of Mathematical Statistics,
  22, 592

\bibitem[\protect\citeauthoryear{Clementel, Madura, Kruip, Icke  \&
  Gull}{Clementel et~al.}{2014}]{2014MNRAS.443.2475C}
Clementel N.,  Madura T.~I.,  Kruip C. J.~H.,  Icke V.,   Gull T.~R.,  2014,
  Monthly Notices of the Royal Astronomical Society, 443, 2475

\bibitem[\protect\citeauthoryear{Clementel, Madura, Kruip, Paardekooper  \&
  Gull}{Clementel et~al.}{2015a}]{2015MNRAS.447.2445C}
Clementel N.,  Madura T.~I.,  Kruip C. J.~H.,  Paardekooper J.~P.,   Gull
  T.~R.,  2015a, Monthly Notices of the Royal Astronomical Society, 447, 2445

\bibitem[\protect\citeauthoryear{Clementel, Madura, Kruip  \&
  Paardekooper}{Clementel et~al.}{2015b}]{2015MNRAS.450.1388C}
Clementel N.,  Madura T.~I.,  Kruip C. J.~H.,   Paardekooper J.~P.,  2015b,
  Monthly Notices of the Royal Astronomical Society, 450, 1388

\bibitem[\protect\citeauthoryear{Conover}{Conover}{1971}]{Conover:1971wa}
Conover W.~J.,  1971, {Practical nonparametric statistics}.
 Vol. 15, New York, Wiley

\bibitem[\protect\citeauthoryear{Conti}{Conti}{1984}]{1984IAUS..105..233C}
Conti P.~S.,  1984, Observational Tests of the Stellar Evolution Theory.
  International Astronomical Union Symposium No. 105, 105, 233

\bibitem[\protect\citeauthoryear{Corcoran, Ishibashi, Swank  \& Petre}{Corcoran
  et~al.}{2001}]{2001ApJ...547.1034C}
Corcoran M.~F.,  Ishibashi K.,  Swank J.~H.,   Petre R.,  2001, The
  Astrophysical Journal, 547, 1034

\bibitem[\protect\citeauthoryear{Corcoran, Hamaguchi, Pittard, Russell, Owocki,
  Parkin  \& Okazaki}{Corcoran et~al.}{2010}]{2010ApJ...725.1528C}
Corcoran M.~F.,  Hamaguchi K.,  Pittard J.~M.,  Russell C. M.~P.,  Owocki
  S.~P.,  Parkin E.~R.,   Okazaki A.,  2010, The Astrophysical Journal, 725,
  1528

\bibitem[\protect\citeauthoryear{Damineli}{Damineli}{1996}]{1996ApJ...460L..49D}
Damineli A.,  1996, Astrophysical Journal Letters v.460, 460, L49

\bibitem[\protect\citeauthoryear{Damineli, Conti  \& Lopes}{Damineli
  et~al.}{1997}]{1997NewA....2..107D}
Damineli A.,  Conti P.~S.,   Lopes D.~F.,  1997, New Astronomy, 2, 107

\bibitem[\protect\citeauthoryear{Damineli, Stahl, Kaufer, Wolf, Quast  \&
  Lopes}{Damineli et~al.}{1998}]{1998yCat..41330299D}
Damineli A.,  Stahl O.,  Kaufer A.,  Wolf B.,  Quast G.,   Lopes D.~F.,  1998,
  VizieR On-line Data Catalog, 413, 30299

\bibitem[\protect\citeauthoryear{Damineli, Lopes  \& Conti}{Damineli
  et~al.}{1999}]{1999ASPC..179..288D}
Damineli A.,  Lopes D.~F.,   Conti P.~S.,  1999, in Eta Carinae at The
  Millennium. p.~288

\bibitem[\protect\citeauthoryear{Damineli et~al.,}{Damineli
  et~al.}{2008a}]{2008MNRAS.384.1649D}
Damineli A.,  et~al., 2008a, Monthly Notices of the Royal Astronomical Society,
  384, 1649

\bibitem[\protect\citeauthoryear{Damineli et~al.,}{Damineli
  et~al.}{2008b}]{2008MNRAS.386.2330D}
Damineli A.,  et~al., 2008b, Monthly Notices of the Royal Astronomical Society,
  386, 2330

\bibitem[\protect\citeauthoryear{Davidson}{Davidson}{1997}]{1997NewA....2..387D}
Davidson K.,  1997, New Astronomy, 2, 387

\bibitem[\protect\citeauthoryear{Davidson}{Davidson}{2002}]{2002ASPC..262..267D}
Davidson K.,  2002, in The High Energy Universe at Sharp Focus: Chandra
  Science. p.~267

\bibitem[\protect\citeauthoryear{Davidson \& Humphreys}{Davidson \&
  Humphreys}{1997}]{1997ARA&A..35....1D}
Davidson K.,  Humphreys R.~M.,  1997, Annual Review of Astronomy and
  Astrophysics, 35, 1

\bibitem[\protect\citeauthoryear{Davidson, Ebbets, Weigelt, Humphreys, Hajian,
  Walborn  \& Rosa}{Davidson et~al.}{1995}]{1995AJ....109.1784D}
Davidson K.,  Ebbets D.,  Weigelt G.,  Humphreys R.~M.,  Hajian A.~R.,  Walborn
  N.~R.,   Rosa M.,  1995, Astronomical Journal (ISSN 0004-6256), 109, 1784

\bibitem[\protect\citeauthoryear{Davidson, Gull, Walborn  \& Gull}{Davidson
  et~al.}{2002}]{2002ecrl.confP...7G}
Davidson K.,  Gull T.,  Walborn N.~R.,   Gull T.~R.,  2002, in Eta Carinae:
  Reading the Legend. p.~7P

\bibitem[\protect\citeauthoryear{Davidson, Mehner, Humphreys, Martin  \&
  Ishibashi}{Davidson et~al.}{2015}]{Davidson:2015gs}
Davidson K.,  Mehner A.,  Humphreys R.~M.,  Martin J.~C.,   Ishibashi K.,
  2015, Astrophysical Journal, 801, L15

\bibitem[\protect\citeauthoryear{Durbin}{Durbin}{1973}]{Durbin:1973tp}
Durbin J. J.~.,  1973

\bibitem[\protect\citeauthoryear{Fahed et~al.,}{Fahed
  et~al.}{2011}]{Fahed:2011gh}
Fahed R.,  et~al., 2011, Monthly Notices of the Royal Astronomical Society,
  418, 2

\bibitem[\protect\citeauthoryear{Falceta-Gon{\c c}alves, Jatenco-Pereira  \&
  Abraham}{Falceta-Gon{\c c}alves et~al.}{2005}]{2005MNRAS.357..895F}
Falceta-Gon{\c c}alves D.,  Jatenco-Pereira V.,   Abraham Z.,  2005, 357, 895

\bibitem[\protect\citeauthoryear{Fisher}{Fisher}{1925}]{Fisher:1925vv}
Fisher R.~A.,  1925, {Statistical methods for research workers}.
Edinburgh, London, Oliver and Boyd

\bibitem[\protect\citeauthoryear{Fisher}{Fisher}{1935}]{Fisher:1935uc}
Fisher R.~A.,  1935, {The design of experiments}.
Edinburgh, London, Oliver and Boyde

\bibitem[\protect\citeauthoryear{Fuhrmann, Chini, Hoffmeister, Lemke, Murphy,
  Seifert  \& Stahl}{Fuhrmann et~al.}{2011}]{2011MNRAS.411.2311F}
Fuhrmann K.,  Chini R.,  Hoffmeister V.~H.,  Lemke R.,  Murphy M.,  Seifert W.,
    Stahl O.,  2011, Monthly Notices of the Royal Astronomical Society, 411,
  2311

\bibitem[\protect\citeauthoryear{Gaviola}{Gaviola}{1950}]{1950ApJ...111..408G}
Gaviola E.,  1950, Astrophysical Journal, 111, 408

\bibitem[\protect\citeauthoryear{Gaviola}{Gaviola}{1953}]{1953ApJ...118..234G}
Gaviola E.,  1953, Astrophysical Journal, 118, 234

\bibitem[\protect\citeauthoryear{Groh \& Damineli}{Groh \&
  Damineli}{2004}]{2004IBVS.5492....1G}
Groh J.~H.,  Damineli A.,  2004, Information Bulletin on Variable Stars, 5492,
  1

\bibitem[\protect\citeauthoryear{Groh et~al.,}{Groh
  et~al.}{2010}]{2010A&A...517A...9G}
Groh J.~H.,  et~al., 2010, Astronomy and Astrophysics, 517, 9

\bibitem[\protect\citeauthoryear{Groh, Hillier, Madura  \& Weigelt}{Groh
  et~al.}{2012a}]{2012MNRAS.423.1623G}
Groh J.~H.,  Hillier D.~J.,  Madura T.~I.,   Weigelt G.,  2012a, Monthly
  Notices of the Royal Astronomical Society, 423, 1623

\bibitem[\protect\citeauthoryear{Groh, Madura, Hillier, Kruip, Kruip  \&
  Weigelt}{Groh et~al.}{2012b}]{2012ApJ...759L...2G}
Groh J.~H.,  Madura T.~I.,  Hillier D.~J.,  Kruip C. J.~H.,  Kruip C. J.~H.,
  Weigelt G.,  2012b, The Astrophysical Journal Letters, 759, L2

\bibitem[\protect\citeauthoryear{Gull, Madura, Groh  \& Corcoran}{Gull
  et~al.}{2011}]{2011ApJ...743L...3G}
Gull T.~R.,  Madura T.~I.,  Groh J.~H.,   Corcoran M.~F.,  2011, The
  Astrophysical Journal Letters, 743, L3

\bibitem[\protect\citeauthoryear{Hamaguchi et~al.,}{Hamaguchi
  et~al.}{2007}]{2007ApJ...663..522H}
Hamaguchi K.,  et~al., 2007, Astrophysical Journal, 663, 522

\bibitem[\protect\citeauthoryear{Hamaguchi et~al.,}{Hamaguchi
  et~al.}{2014}]{2014ApJ...784..125H}
Hamaguchi K.,  et~al., 2014, The Astrophysical Journal, 784, 125

\bibitem[\protect\citeauthoryear{Hearnshaw, Barnes, Kershaw, Frost, Graham,
  Ritchie  \& Nankivell}{Hearnshaw et~al.}{2002}]{2002ExA....13...59H}
Hearnshaw J.~B.,  Barnes S.~I.,  Kershaw G.~M.,  Frost N.,  Graham G.,  Ritchie
  R.,   Nankivell G.~R.,  2002, Experimental Astronomy, 13, 59

\bibitem[\protect\citeauthoryear{Henley, Corcoran, Pittard, Stevens, Hamaguchi
  \& Gull}{Henley et~al.}{2008}]{2008ApJ...680..705H}
Henley D.~B.,  Corcoran M.~F.,  Pittard J.~M.,  Stevens I.~R.,  Hamaguchi K.,
  Gull T.~R.,  2008, The Astrophysical Journal, 680, 705

\bibitem[\protect\citeauthoryear{Hillier, Davidson, Ishibashi  \& Gull}{Hillier
  et~al.}{2001}]{2001ApJ...553..837H}
Hillier D.~J.,  Davidson K.,  Ishibashi K.,   Gull T.,  2001, The Astrophysical
  Journal, 553, 837

\bibitem[\protect\citeauthoryear{Hillier et~al.,}{Hillier
  et~al.}{2006}]{2006ApJ...642.1098H}
Hillier D.~J.,  et~al., 2006, The Astrophysical Journal, 642, 1098

\bibitem[\protect\citeauthoryear{Hollander \& Wolfe}{Hollander \&
  Wolfe}{1973}]{Hollander:1973wz}
Hollander M.,  Wolfe D. A. j.~a.,  1973, {Nonparametric statistical methods}.
 Vol. 17, New York, Wiley

\bibitem[\protect\citeauthoryear{Humphreys}{Humphreys}{1978}]{1978ApJ...219..445H}
Humphreys R.~M.,  1978, Astrophysical Journal, 219, 445

\bibitem[\protect\citeauthoryear{Humphreys \& {HST-FOS eta Car Team}}{Humphreys
  \& {HST-FOS eta Car Team}}{1999}]{1999ASPC..179..107H}
Humphreys R.~M.,  {HST-FOS eta Car Team} 1999, Eta Carinae At The Millennium,
  179, 107

\bibitem[\protect\citeauthoryear{Kashi \& Soker}{Kashi \&
  Soker}{2008}]{2008MNRAS.390.1751K}
Kashi A.,  Soker N.,  2008, Monthly Notices of the Royal Astronomical Society,
  390, 1751

\bibitem[\protect\citeauthoryear{Kashi \& Soker}{Kashi \&
  Soker}{2009}]{2009MNRAS.397.1426K}
Kashi A.,  Soker N.,  2009, Publications of the Astron. Soc. of Australia, 397,
  1426

\bibitem[\protect\citeauthoryear{Kashi \& Soker}{Kashi \&
  Soker}{2015}]{Kashi:2015ul}
Kashi A.,  Soker N.,  2015, eprint arXiv:1508.03576

\bibitem[\protect\citeauthoryear{Knuth}{Knuth}{2011}]{Knuth:2011fq}
Knuth D.~E.,  2011, Choice Reviews Online, 48, 48

\bibitem[\protect\citeauthoryear{Madura \& Groh}{Madura \&
  Groh}{2012}]{2012ApJ...746L..18M}
Madura T.~I.,  Groh J.~H.,  2012, The Astrophysical Journal Letters, 746, L18

\bibitem[\protect\citeauthoryear{Madura \& Owocki}{Madura \&
  Owocki}{2010}]{2010RMxAC..38...52M}
Madura T.~I.,  Owocki S.~P.,  2010, The Interferometric View on Hot Stars (Eds.
  Th. Rivinius {\&} M. Cur{\'e}) Revista Mexicana de Astronom{\'\i}a y
  Astrof{\'\i}sica (Serie de Conferencias) Vol. 38, 38, 52

\bibitem[\protect\citeauthoryear{Madura, Gull, Owocki, Groh, Okazaki  \&
  Russell}{Madura et~al.}{2012}]{2012MNRAS.420.2064M}
Madura T.~I.,  Gull T.~R.,  Owocki S.~P.,  Groh J.~H.,  Okazaki A.~T.,
  Russell C. M.~P.,  2012, Monthly Notices of the Royal Astronomical Society,
  420, 2064

\bibitem[\protect\citeauthoryear{Madura et~al.,}{Madura
  et~al.}{2013}]{2013MNRAS.436.3820M}
Madura T.~I.,  et~al., 2013, Monthly Notices of the Royal Astronomical Society,
  436, 3820

\bibitem[\protect\citeauthoryear{Marsaglia, Tsang  \& Wang}{Marsaglia
  et~al.}{2003}]{Marsaglia:2003vb}
Marsaglia G.,  Tsang W.~W.,   Wang J.,  2003, Journal of Statistical Software,
  8, 1

\bibitem[\protect\citeauthoryear{Martin, Davidson, Humphreys, Hillier  \&
  Ishibashi}{Martin et~al.}{2006}]{2006ApJ...640..474M}
Martin J.~C.,  Davidson K.,  Humphreys R.~M.,  Hillier D.~J.,   Ishibashi K.,
  2006, The Astrophysical Journal, 640, 474

\bibitem[\protect\citeauthoryear{Mehner, Davidson, Ferland  \&
  Humphreys}{Mehner et~al.}{2010}]{2010ApJ...710..729M}
Mehner A.,  Davidson K.,  Ferland G.~J.,   Humphreys R.~M.,  2010, The
  Astrophysical Journal, 710, 729

\bibitem[\protect\citeauthoryear{Mehner, Davidson, Martin, Humphreys, Ishibashi
   \& Ferland}{Mehner et~al.}{2011}]{2011ApJ...740...80M}
Mehner A.,  Davidson K.,  Martin J.~C.,  Humphreys R.~M.,  Ishibashi K.,
  Ferland G.~J.,  2011, The Astrophysical Journal, 740, 80

\bibitem[\protect\citeauthoryear{Mehner et~al.,}{Mehner
  et~al.}{2015}]{2015arXiv150404940M}
Mehner A.,  et~al., 2015, Astronomy and Astrophysics, 578, A122

\bibitem[\protect\citeauthoryear{Moffat \& Corcoran}{Moffat \&
  Corcoran}{2009}]{2009ApJ...707..693M}
Moffat A. F.~J.,  Corcoran M.~F.,  2009, The Astrophysical Journal, 707, 693

\bibitem[\protect\citeauthoryear{Nielsen, Corcoran, Gull, Hillier, Hamaguchi,
  Ivarsson  \& Lindler}{Nielsen et~al.}{2007}]{2007ApJ...660..669N}
Nielsen K.~E.,  Corcoran M.~F.,  Gull T.~R.,  Hillier D.~J.,  Hamaguchi K.,
  Ivarsson S.,   Lindler D.~J.,  2007, The Astrophysical Journal, 660, 669

\bibitem[\protect\citeauthoryear{Okazaki, Owocki, Russell  \& Corcoran}{Okazaki
  et~al.}{2008}]{2008MNRAS.388L..39O}
Okazaki A.~T.,  Owocki S.~P.,  Russell C. M.~P.,   Corcoran M.~F.,  2008,
  Monthly Notices of the Royal Astronomical Society: Letters, 388, L39

\bibitem[\protect\citeauthoryear{Parkin, Pittard, Corcoran, Hamaguchi  \&
  Stevens}{Parkin et~al.}{2009}]{2009MNRAS.394.1758P}
Parkin E.~R.,  Pittard J.~M.,  Corcoran M.~F.,  Hamaguchi K.,   Stevens I.~R.,
  2009, Monthly Notices of the Royal Astronomical Society, 394, 1758

\bibitem[\protect\citeauthoryear{Parkin, Pittard, Corcoran  \&
  Hamaguchi}{Parkin et~al.}{2011}]{2011ApJ...726..105P}
Parkin E.~R.,  Pittard J.~M.,  Corcoran M.~F.,   Hamaguchi K.,  2011, The
  Astrophysical Journal, 726, 105

\bibitem[\protect\citeauthoryear{Pittard \& Corcoran}{Pittard \&
  Corcoran}{2002}]{2002A&A...383..636P}
Pittard J.~M.,  Corcoran M.~F.,  2002, Astronomy and Astrophysics, 383, 636

\bibitem[\protect\citeauthoryear{Plavchan, Jura, Kirkpatrick, Cutri  \&
  Gallagher}{Plavchan et~al.}{2008}]{2008ApJS..175..191P}
Plavchan P.,  Jura M.,  Kirkpatrick J.~D.,  Cutri R.~M.,   Gallagher S.~C.,
  2008, The Astrophysical Journal Supplement Series, 175, 191

\bibitem[\protect\citeauthoryear{{R Core Team}}{{R Core
  Team}}{2014}]{Team:2014wf}
{R Core Team} 2014, {R: A Language and Environment for Statistical Computing}.
R Foundation for Statistical Computing, Vienna, Austria

\bibitem[\protect\citeauthoryear{Richardson, Gies, Henry,
  Fern{\'a}ndez-Laj{\'u}s  \& Okazaki}{Richardson
  et~al.}{2010}]{2010AJ....139.1534R}
Richardson N.~D.,  Gies D.~R.,  Henry T.~J.,  Fern{\'a}ndez-Laj{\'u}s E.,
  Okazaki A.~T.,  2010, The Astronomical Journal, 139, 1534

\bibitem[\protect\citeauthoryear{Richardson, Gies, Gull, Moffat  \&
  St-Jean}{Richardson et~al.}{2015}]{Richardson:2015ur}
Richardson N.~D.,  Gies D.~R.,  Gull T.~R.,  Moffat A. F.~J.,   St-Jean L.,
  2015, The Astronomical Journal, 150, 109

\bibitem[\protect\citeauthoryear{Rivinius, Wolf, Stahl  \& Kaufer}{Rivinius
  et~al.}{2001}]{2001ASPC..242...29R}
Rivinius T.,  Wolf B.,  Stahl O.,   Kaufer A.,  2001, in Eta Carinae and Other
  Mysterious Stars: The Hidden Opportunities of Emission Spectroscopy. p.~29

\bibitem[\protect\citeauthoryear{Russell}{Russell}{2013}]{2013PhDT.......164R}
Russell C. M.~P.,  2013, PhD thesis, ProQuest Dissertations And Theses; Thesis
  (Ph.D.)--University of Delaware, University of Delaware

\bibitem[\protect\citeauthoryear{Scholz \& Stephens}{Scholz \&
  Stephens}{1987}]{Scholz:1987eq}
Scholz F.~W.,  Stephens M.~A.,  1987, Journal of the American Statistical
  Association, 82, 918

\bibitem[\protect\citeauthoryear{Smith}{Smith}{2006}]{2006ApJ...644.1151S}
Smith N.,  2006, The Astrophysical Journal, 644, 1151

\bibitem[\protect\citeauthoryear{Smith \& Gehrz}{Smith \&
  Gehrz}{1998}]{1998AJ....116..823S}
Smith N.,  Gehrz R.~D.,  1998, The Astronomical Journal, 116, 823

\bibitem[\protect\citeauthoryear{Smith, Gehrz, Hinz, Hoffmann, Hora, Mamajek
  \& Meyer}{Smith et~al.}{2003a}]{2003AJ....125.1458S}
Smith N.,  Gehrz R.~D.,  Hinz P.~M.,  Hoffmann W.~F.,  Hora J.~L.,  Mamajek
  E.~E.,   Meyer M.~R.,  2003a, The Astronomical Journal, 125, 1458

\bibitem[\protect\citeauthoryear{Smith, Davidson, Gull, Ishibashi  \&
  Hillier}{Smith et~al.}{2003b}]{2003ApJ...586..432S}
Smith N.,  Davidson K.,  Gull T.~R.,  Ishibashi K.,   Hillier D.~J.,  2003b,
  The Astrophysical Journal, 586, 432

\bibitem[\protect\citeauthoryear{Stahl, Weis, Bomans, Davidson, Gull  \&
  Humphreys}{Stahl et~al.}{2005}]{2005A&A...435..303S}
Stahl O.,  Weis K.,  Bomans D.~J.,  Davidson K.,  Gull T.~R.,   Humphreys
  R.~M.,  2005, Astronomy and Astrophysics, 435, 303

\bibitem[\protect\citeauthoryear{Steffen et~al.,}{Steffen
  et~al.}{2014}]{2014MNRAS.442.3316S}
Steffen W.,  et~al., 2014, Monthly Notices of the Royal Astronomical Society,
  442, 3316

\bibitem[\protect\citeauthoryear{Steiner \& Damineli}{Steiner \&
  Damineli}{2004}]{2004ApJ...612L.133S}
Steiner J.~E.,  Damineli A.,  2004, The Astrophysical Journal, 612, L133

\bibitem[\protect\citeauthoryear{Stellingwerf}{Stellingwerf}{1978}]{1978ApJ...224..953S}
Stellingwerf R.~F.,  1978, Astrophysical Journal, 224, 953

\bibitem[\protect\citeauthoryear{Teodoro, Damineli, Sharp, Groh  \&
  Barbosa}{Teodoro et~al.}{2008}]{2008MNRAS.387..564T}
Teodoro M.,  Damineli A.,  Sharp R.~G.,  Groh J.~H.,   Barbosa C.~L.,  2008,
  Monthly Notices of the Royal Astronomical Society, 387, 564

\bibitem[\protect\citeauthoryear{Teodoro et~al.,}{Teodoro
  et~al.}{2012}]{2012ApJ...746...73T}
Teodoro M.,  et~al., 2012, The Astrophysical Journal, 746, 73

\bibitem[\protect\citeauthoryear{Teodoro, Madura, Gull, Corcoran  \&
  Hamaguchi}{Teodoro et~al.}{2013}]{2013ApJ...773L..16T}
Teodoro M.,  Madura T.~I.,  Gull T.~R.,  Corcoran M.~F.,   Hamaguchi K.,  2013,
  The Astrophysical Journal Letters, 773, L16

\bibitem[\protect\citeauthoryear{Tokovinin, Fischer, Bonati, Giguere, Moore,
  Schwab, Spronck  \& Szymkowiak}{Tokovinin et~al.}{2013}]{2013PASP..125.1336T}
Tokovinin A.,  Fischer D.~A.,  Bonati M.,  Giguere M.~J.,  Moore P.,  Schwab
  C.,  Spronck J. F.~P.,   Szymkowiak A.,  2013, Publications of the
  Astronomical Society of the Pacific, 125, 1336

\bibitem[\protect\citeauthoryear{Verner, Bruhweiler  \& Gull}{Verner
  et~al.}{2005}]{2005ApJ...624..973V}
Verner E.,  Bruhweiler F.,   Gull T.,  2005, The Astrophysical Journal, 624,
  973

\bibitem[\protect\citeauthoryear{Zanella, Wolf  \& Stahl}{Zanella
  et~al.}{1984}]{1984A&A...137...79Z}
Zanella R.,  Wolf B.,   Stahl O.,  1984, Astronomy and Astrophysics (ISSN
  0004-6361), 137, 79

\bibitem[\protect\citeauthoryear{Zethson, Zethson, Johansson, Davidson,
  Humphreys, Ishibashi  \& Ebbets}{Zethson et~al.}{1999}]{1999A&A...344..211Z}
Zethson T.,  Zethson T.,  Johansson S.,  Davidson K.,  Humphreys R.~M.,
  Ishibashi K.,   Ebbets D.,  1999, Astronomy and Astrophysics, 344, 211

\makeatother
\end{thebibliography}

\appendix
\Cref{tab:obs} lists all the \btxt{2014.6} measurements presented in this paper, as well as information about the quality and wavelength coverage of each spectrum used.

\LongTables
\begin{deluxetable*}{ccccccc}

\tabletypesize{\footnotesize}
\tablecolumns{7}
\tablewidth{0pt}

\tablecaption{\btxt{\heii~emission line measurements and spectral characteristics of the data obtained during the international campaign to monitor the 2014.6 periastron passage.}\label{tab:obs}}

\tablehead{
                                              & \multirow{3}{*}{$EW$} & Velocity of  &  &  & Wavelength & \\ 
\multirow{3}{*}{JD-2450000} &  & the peak & $\Delta\mathrm{v\tablenotemark{a}}$ & \multirow{3}{*}{\textit{SNR}\tablenotemark{b}} & coverage & \multirow{3}{*}{Observatory} \\ 
                                              & (\AA) & (km\,s$^{-1}$) & (km\,s$^{-1}$) &  & (\AA) &  \\
 }

\startdata
4967.47314 & $+0.16 \pm 0.08$ & $  -10$ & 185.7 &   1579 & 3499--6168 & SOAR \\
4978.45698 & $+0.16 \pm 0.08$ & $  -28$ & 185.8 &   1079 & 3497--6167 & SOAR \\
5012.50000 & $+0.03 \pm 0.03$ & $ -678$ & 39.5 &    955 & 4560--4826 & HST \\
5171.50000 & $+0.20 \pm 0.03$ & $ -678$ & 39.5 &    629 & 4560--4826 & HST \\
5495.50000 & $+0.35 \pm 0.03$ & $ -678$ & 39.5 &    334 & 4560--4826 & HST \\
5689.50746 & $-0.23 \pm 0.08$ & $  -23$ & 92.5 &    350 & 3509--4839 & SOAR \\
5729.53926 & $+0.06 \pm 0.08$ & $ +522$ & 92.1 &    342 & 3516--4841 & SOAR \\
5768.49673 & $+0.12 \pm 0.08$ & $ +514$ & 92.0 &    353 & 3517--4840 & SOAR \\
5856.82809 & $+0.08 \pm 0.08$ & $ +526$ & 92.0 &    307 & 3517--4839 & SOAR \\
5885.50000 & $+0.16 \pm 0.03$ & $ -678$ & 39.5 &    272 & 4560--4826 & HST \\
5893.79688 & $-0.24 \pm 0.08$ & $ -668$ & 92.1 &    396 & 3509--4832 & SOAR \\
5930.77193 & $-0.06 \pm 0.08$ & $  -59$ & 92.0 &    379 & 3513--4836 & SOAR \\
5955.81938 & $+0.14 \pm 0.08$ & $ -672$ & 92.0 &    250 & 3518--4841 & SOAR \\
5982.61373 & $+0.11 \pm 0.08$ & $ +526$ & 91.9 &    346 & 3517--4839 & SOAR \\
5989.68923 & $+0.01 \pm 0.13$ & $ +530$ & 3.5 &    640 & 4505--4755 & CTIO \\
5993.75221 & $+0.04 \pm 0.13$ & $ +530$ & 3.5 &    620 & 4505--4755 & CTIO \\
6001.72621 & $+0.08 \pm 0.13$ & $ +530$ & 3.5 &    678 & 4505--4755 & CTIO \\
6003.63393 & $+0.05 \pm 0.08$ & $ +514$ & 92.0 &    287 & 3518--4840 & SOAR \\
6013.68632 & $+0.22 \pm 0.13$ & $ -684$ & 3.5 &    594 & 4505--4755 & CTIO \\
6039.52767 & $+0.02 \pm 0.08$ & $ +525$ & 92.2 &    422 & 3511--4836 & SOAR \\
6079.49884 & $+0.03 \pm 0.08$ & $ +515$ & 92.1 &    299 & 3519--4843 & SOAR \\
6116.49641 & $-0.13 \pm 0.08$ & $ +530$ & 92.0 &    279 & 3521--4844 & SOAR \\
6136.46326 & $+0.08 \pm 0.08$ & $ +527$ & 92.2 &    289 & 3515--4841 & SOAR \\
6218.50000 & $+0.22 \pm 0.03$ & $ -358$ & 39.5 &    368 & 4560--4826 & HST \\
6221.89057 & $+0.20 \pm 0.13$ & $ +530$ & 3.5 &    584 & 4505--4755 & CTIO \\
6236.84632 & $-0.06 \pm 0.13$ & $ -684$ & 3.5 &    548 & 4505--4755 & CTIO \\
6238.87015 & $-0.04 \pm 0.13$ & $ -684$ & 3.5 &    410 & 4505--4755 & CTIO \\
6246.79354 & $+0.01 \pm 0.08$ & $ +521$ & 92.0 &    323 & 3517--4839 & SOAR \\
6248.81477 & $-0.10 \pm 0.13$ & $ +530$ & 3.5 &    606 & 4505--4755 & CTIO \\
6254.88181 & $-0.19 \pm 0.13$ & $ -684$ & 3.5 &    558 & 4505--4755 & CTIO \\
6260.78284 & $-0.05 \pm 0.13$ & $ +530$ & 3.5 &    433 & 4505--4755 & CTIO \\
6275.81899 & $+0.28 \pm 0.13$ & $ -684$ & 3.5 &    673 & 4505--4755 & CTIO \\
6278.81225 & $+0.26 \pm 0.08$ & $ +518$ & 92.0 &    333 & 3516--4839 & SOAR \\
6289.79792 & $+0.16 \pm 0.13$ & $ -684$ & 3.5 &    788 & 4505--4755 & CTIO \\
6346.69729 & $+0.09 \pm 0.08$ & $ -673$ & 91.9 &    439 & 3520--4841 & SOAR \\
6361.58149 & $-0.24 \pm 0.08$ & $ +526$ & 91.9 &    327 & 3518--4839 & SOAR \\
6361.64473 & $+0.12 \pm 0.13$ & $ +530$ & 3.1 &    742 & 4505--4755 & CTIO \\
6401.58757 & $-0.07 \pm 0.13$ & $ +530$ & 3.1 &    772 & 4505--4755 & CTIO \\
6417.58964 & $-0.09 \pm 0.13$ & $ +530$ & 3.1 &    675 & 4505--4755 & CTIO \\
6428.48444 & $-0.03 \pm 0.08$ & $ +522$ & 92.3 &    460 & 3511--4837 & SOAR \\
6428.64093 & $-0.03 \pm 0.13$ & $ +131$ & 3.8 &   2046 & 4575--4800 & CTIO \\
6452.52706 & $-0.05 \pm 0.08$ & $  +31$ & 92.2 &    357 & 3514--4840 & SOAR \\
6488.47634 & $-0.19 \pm 0.08$ & $  +14$ & 91.9 &    484 & 3516--4838 & SOAR \\
6538.50000 & $+0.10 \pm 0.03$ & $  -17$ & 39.5 &    265 & 4560--4826 & HST \\
6607.84573 & $+0.12 \pm 0.13$ & $ +530$ & 3.1 &    497 & 4505--4755 & CTIO \\
6612.86785 & $-0.22 \pm 0.13$ & $ +324$ & 3.1 &    577 & 4505--4755 & CTIO \\
6613.83605 & $-0.34 \pm 0.08$ & $  +15$ & 91.1 &    420 & 3632--4941 & SOAR \\
6655.76290 & $+0.01 \pm 0.15$ & $   -1$ & 47.2 &    615 & 6398--6720 & SOAR \\
6656.70393 & $-0.09 \pm 0.08$ & $  -30$ & 3.1 &   1225 & 4283--4961 & CTIO \\
6659.70988 & $-0.08 \pm 0.13$ & $  -29$ & 3.1 &    690 & 4505--4755 & CTIO \\
6664.67479 & $-0.17 \pm 0.13$ & $ +530$ & 3.1 &    607 & 4505--4755 & CTIO \\
6670.78253 & $-0.01 \pm 0.13$ & $  -31$ & 3.1 &    657 & 4505--4755 & CTIO \\
6672.83764 & $-0.04 \pm 0.13$ & $  -71$ & 3.1 &    766 & 4505--4755 & CTIO \\
6677.75915 & $-0.17 \pm 0.13$ & $  +12$ & 3.1 &    625 & 4505--4755 & CTIO \\
6687.70761 & $-0.10 \pm 0.13$ & $  -37$ & 3.1 &    755 & 4505--4755 & CTIO \\
6690.69063 & $-0.18 \pm 0.13$ & $   -6$ & 3.1 &    800 & 4505--4755 & CTIO \\
6697.72572 & $-0.27 \pm 0.08$ & $   -2$ & 3.1 &    743 & 4505--4755 & CTIO \\
6698.77338 & $-0.17 \pm 0.08$ & $  +55$ & 3.7 &   1814 & 4505--4755 & CTIO \\
6701.62899 & $-0.03 \pm 0.13$ & $  +10$ & 99.7 &    399 & 4575--4800 & OPD (Lhires\,{\sc iii}) \\
6701.74483 & $-0.26 \pm 0.03$ & $ +136$ & 46.1 &    184 & 4571--5052 & OPD (Coud\'e) \\
6705.50000 & $-0.18 \pm 0.03$ & $  +22$ & 39.5 &    374 & 4396--4894 & HST \\
6710.66775 & $-0.10 \pm 0.03$ & $  +14$ & 3.1 &    911 & 4560--4826 & CTIO \\
6712.69246 & $-0.23 \pm 0.13$ & $   -7$ & 3.1 &    614 & 4505--4755 & CTIO \\
6718.70977 & $-0.15 \pm 0.13$ & $  -45$ & 3.1 &    680 & 4505--4755 & CTIO \\
6724.59005 & $-0.22 \pm 0.13$ & $  -11$ & 3.7 &   1944 & 4505--4755 & CTIO \\
6725.56888 & $-0.12 \pm 0.13$ & $ -103$ & 3.1 &    700 & 4575--4800 & CTIO \\
6729.55975 & $-0.23 \pm 0.13$ & $ -126$ & 3.0 &    946 & 4505--4755 & CTIO \\
6730.60581 & $-0.23 \pm 0.15$ & $ +527$ & 26.6 &    655 & 6399--6720 & CASLEO \\
6732.60351 & $-0.20 \pm 0.13$ & $ -111$ & 3.0 &    899 & 4505--4755 & CTIO \\
6739.49790 & $-0.22 \pm 0.13$ & $  -42$ & 91.3 &    934 & 4540--4760 & SOAR \\
6739.54616 & $-0.22 \pm 0.13$ & $  -42$ & 3.1 &    828 & 4505--4755 & CTIO \\
6746.51736 & $-0.39 \pm 0.35$ & $  -37$ & 3.1 &    628 & 3627--4940 & CTIO \\
6750.53884 & $-0.30 \pm 0.13$ & $   +2$ & 3.1 &    642 & 4505--4755 & CTIO \\
6754.55102 & $-0.31 \pm 0.13$ & $  -44$ & 3.1 &    702 & 4505--4755 & CTIO \\
6759.52573 & $-0.25 \pm 0.13$ & $  -17$ & 91.4 &    486 & 4505--4755 & SOAR \\
6763.55612 & $-0.33 \pm 0.13$ & $  -36$ & 29.1 &   1082 & 4505--4755 & CASLEO \\
6765.58734 & $-0.40 \pm 0.35$ & $   +7$ & 3.0 &    878 & 3625--4940 & CTIO \\
6766.54334 & $-0.25 \pm 0.13$ & $  -61$ & 3.1 &    987 & 4540--4760 & CTIO \\
6767.50868 & $-0.35 \pm 0.13$ & $  +14$ & 3.1 &    908 & 4505--4755 & CTIO \\
6769.52337 & $-0.23 \pm 0.13$ & $   -1$ & 46.1 &    517 & 4505--4755 & OPD (Coud\'e) \\
6772.43126 & $-0.54 \pm 0.13$ & $  -73$ & 100.1 &    201 & 4505--4755 & OPD (Lhires\,{\sc iii}) \\
6774.54408 & $-0.33 \pm 0.13$ & $ -152$ & 3.1 &    784 & 4461--4959 & CTIO \\
6775.46760 & $-0.53 \pm 0.13$ & $  +28$ & 3.7 &   1968 & 4571--5055 & CTIO \\
6777.53611 & $-0.45 \pm 0.13$ & $  +71$ & 3.7 &   1819 & 4505--4755 & CTIO \\
6781.47143 & $-0.31 \pm 0.13$ & $ -105$ & 3.1 &    911 & 4575--4800 & CTIO \\
6784.84146 & $-0.49 \pm 0.13$ & $  -33$ & 3.2 &   1252 & 4575--4800 & MJUO \\
6786.50087 & $-0.56 \pm 0.13$ & $  -43$ & 3.7 &   2346 & 4505--4755 & CTIO \\
6786.80812 & $-0.78 \pm 0.08$ & $  -43$ & 3.2 &   1041 & 4586--4818 & MJUO \\
6787.50338 & $-0.62 \pm 0.13$ & $  -46$ & 3.7 &   1774 & 4575--4800 & CTIO \\
6787.80222 & $-0.80 \pm 0.08$ & $  -54$ & 3.2 &   1390 & 4586--4817 & MJUO \\
6789.57938 & $-0.80 \pm 0.13$ & $ -135$ & 26.5 &    519 & 4575--4800 & CASLEO \\
6791.55016 & $-0.63 \pm 0.08$ & $ -135$ & 3.1 &   1017 & 4586--4817 & CTIO \\
6793.39493 & $-0.57 \pm 0.13$ & $ -159$ & 100.2 &    585 & 4540--4760 & OPD (Lhires\,{\sc iii}) \\
6793.43162 & $-0.52 \pm 0.13$ & $ -153$ & 100.2 &    260 & 4505--4755 & OPD (Lhires\,{\sc iii}) \\
6793.51309 & $-0.61 \pm 0.13$ & $ -163$ & 26.5 &   1712 & 4569--5053 & CASLEO \\
6795.44909 & $-0.63 \pm 0.13$ & $ -185$ & 100.3 &    114 & 4569--5053 & OPD (Lhires\,{\sc iii}) \\
6795.44909 & $-0.63 \pm 0.13$ & $ -185$ & 100.3 &    114 & 4540--4760 & OPD (Lhires\,{\sc iii}) \\
6795.50702 & $-0.72 \pm 0.13$ & $ -174$ & 3.1 &    971 & 4570--5055 & CTIO \\
6796.46794 & $-0.52 \pm 0.13$ & $ -175$ & 93.7 &   1262 & 4570--5055 & SOAR \\
6796.66742 & $-0.72 \pm 0.13$ & $ -158$ & 100.3 &    103 & 4505--4755 & OPD (Lhires\,{\sc iii}) \\
6797.51150 & $-0.80 \pm 0.35$ & $ -159$ & 100.3 &    564 & 3593--4940 & OPD (Lhires\,{\sc iii}) \\
6797.56402 & $-0.69 \pm 0.13$ & $ -174$ & 26.5 &    462 & 4565--5049 & CASLEO \\
6800.41794 & $-0.87 \pm 0.13$ & $  -93$ & 46.9 &    433 & 4556--5040 & OPD (Coud\'e) \\
6800.46206 & $-0.65 \pm 0.13$ & $ -102$ & 3.1 &    838 & 4540--4760 & CTIO \\
6801.49770 & $-0.64 \pm 0.13$ & $  -94$ & 3.1 &    737 & 4035--5176 & CTIO \\
6802.50447 & $-0.79 \pm 0.13$ & $   +3$ & 3.7 &   1702 & 4505--4755 & CTIO \\
6803.52719 & $-0.70 \pm 0.13$ & $ -100$ & 3.1 &   1034 & 4505--4755 & CTIO \\
6805.46847 & $-0.93 \pm 0.13$ & $  -27$ & 100.3 &    388 & 4575--4800 & OPD (Lhires\,{\sc iii}) \\
6805.48993 & $-1.01 \pm 0.13$ & $   -4$ & 100.3 &    275 & 4505--4755 & OPD (Lhires\,{\sc iii}) \\
6806.52963 & $-0.84 \pm 0.13$ & $  -46$ & 3.7 &    460 & 4568--5052 & CTIO \\
6807.61259 & $-0.85 \pm 0.13$ & $  +68$ & 3.7 &    490 & 4568--5052 & CTIO \\
6809.45843 & $-0.79 \pm 0.13$ & $   -4$ & 3.7 &   2494 & 4575--4800 & CTIO \\
6810.48189 & $-0.69 \pm 0.13$ & $  -94$ & 3.1 &    868 & 4575--4800 & CTIO \\
6811.48702 & $-0.79 \pm 0.13$ & $ -260$ & 3.7 &   2121 & 4575--4800 & CTIO \\
6812.45066 & $-0.60 \pm 0.13$ & $ -166$ & 99.3 &    151 & 4505--4755 & OPD (Lhires\,{\sc iii}) \\
6813.40174 & $-0.76 \pm 0.13$ & $ -239$ & 99.9 &    567 & 4575--4800 & OPD (Lhires\,{\sc iii}) \\
6813.42771 & $-0.89 \pm 0.13$ & $ -230$ & 99.9 &    193 & 4574--5054 & OPD (Lhires\,{\sc iii}) \\
6814.41074 & $-0.66 \pm 0.13$ & $ -176$ & 99.9 &    732 & 4573--5056 & OPD (Lhires\,{\sc iii}) \\
6814.43755 & $-0.62 \pm 0.13$ & $ -165$ & 99.9 &    186 & 4573--5056 & OPD (Lhires\,{\sc iii}) \\
6816.43650 & $-0.73 \pm 0.13$ & $ -205$ & 99.8 &    393 & 4573--5055 & OPD (Lhires\,{\sc iii}) \\
6816.43650 & $-0.77 \pm 0.13$ & $ -207$ & 99.8 &    280 & 4573--5055 & OPD (Lhires\,{\sc iii}) \\
6817.50000 & $-0.94 \pm 0.13$ & $ -178$ & 39.5 &    359 & 4573--5055 & HST \\
6818.42255 & $-0.91 \pm 0.13$ & $ -174$ & 99.8 &    233 & 4573--5055 & OPD (Lhires\,{\sc iii}) \\
6818.50425 & $-0.67 \pm 0.03$ & $ -185$ & 3.1 &    888 & 4560--4826 & CTIO \\
6820.41744 & $-0.89 \pm 0.13$ & $ -148$ & 102.4 &    546 & 4587--5069 & OPD (Lhires\,{\sc iii}) \\
6821.80127 & $-1.09 \pm 0.13$ & $  -67$ & 3.2 &    901 & 4505--4755 & MJUO \\
6822.38703 & $-1.14 \pm 0.13$ & $ -136$ & 102.4 &    525 & 4577--5055 & OPD (Lhires\,{\sc iii}) \\
6822.45982 & $-1.32 \pm 0.22$ & $ -106$ & 91.2 &   2197 & 4586--4818 & SOAR \\
6823.45482 & $-1.20 \pm 0.13$ & $  -58$ & 99.9 &    522 & 4577--5055 & OPD (Lhires\,{\sc iii}) \\
6823.53806 & $-1.12 \pm 0.35$ & $  -68$ & 3.1 &    772 & 3626--4938 & CTIO \\
6823.86265 & $-1.16 \pm 0.13$ & $  -62$ & 3.2 &    960 & 4575--5057 & MJUO \\
6824.51692 & $-1.47 \pm 0.13$ & $  -58$ & 3.1 &   1085 & 4505--4755 & CTIO \\
6825.44362 & $-1.27 \pm 0.22$ & $  -89$ & 99.9 &    719 & 4589--4818 & OPD (Lhires\,{\sc iii}) \\
6825.46968 & $-1.26 \pm 0.13$ & $  -90$ & 99.9 &    379 & 4505--4755 & OPD (Lhires\,{\sc iii}) \\
6825.50553 & $-1.59 \pm 0.13$ & $  -68$ & 3.7 &   2021 & 4574--5057 & CTIO \\
6826.42638 & $-1.30 \pm 0.13$ & $ -104$ & 99.9 &    924 & 4574--5057 & OPD (Lhires\,{\sc iii}) \\
6826.45073 & $-1.30 \pm 0.13$ & $  -83$ & 99.9 &    339 & 4575--4800 & OPD (Lhires\,{\sc iii}) \\
6826.85107 & $-1.37 \pm 0.13$ & $  -74$ & 3.2 &    886 & 4574--5057 & MJUO \\
6827.40873 & $-1.42 \pm 0.13$ & $ -113$ & 99.9 &    835 & 4575--5057 & OPD (Lhires\,{\sc iii}) \\
6827.43762 & $-1.21 \pm 0.22$ & $ -107$ & 99.9 &    622 & 4586--4818 & OPD (Lhires\,{\sc iii}) \\
6827.77608 & $-1.51 \pm 0.13$ & $ -119$ & 3.2 &   1263 & 4574--5057 & MJUO \\
6828.40903 & $-1.38 \pm 0.13$ & $  -67$ & 102.4 &    602 & 4574--5057 & OPD (Lhires\,{\sc iii}) \\
6829.50184 & $-1.57 \pm 0.22$ & $  -42$ & 3.1 &    481 & 4586--4818 & CTIO \\
6830.39501 & $-1.31 \pm 0.13$ & $  -11$ & 99.9 &    637 & 4577--5055 & OPD (Lhires\,{\sc iii}) \\
6830.39501 & $-1.32 \pm 0.13$ & $  -23$ & 99.9 &    660 & 4505--4755 & OPD (Lhires\,{\sc iii}) \\
6831.47240 & $-1.72 \pm 0.13$ & $   -8$ & 91.3 &    411 & 4575--5057 & SOAR \\
6832.50051 & $-1.43 \pm 0.13$ & $  -32$ & 3.1 &    908 & 4575--5057 & CTIO \\
6833.37832 & $-1.15 \pm 0.35$ & $  -63$ & 99.9 &    800 & 3626--4938 & OPD (Lhires\,{\sc iii}) \\
6833.37832 & $-1.20 \pm 0.13$ & $  -60$ & 99.9 &    238 & 4505--4755 & OPD (Lhires\,{\sc iii}) \\
6833.50054 & $-1.34 \pm 0.13$ & $  -38$ & 3.7 &   1941 & 4574--5057 & CTIO \\
6835.36656 & $-1.35 \pm 0.13$ & $ -159$ & 102.4 &    615 & 4574--5057 & OPD (Lhires\,{\sc iii}) \\
6835.53378 & $-1.17 \pm 0.13$ & $ -132$ & 3.1 &    489 & 4575--4800 & CTIO \\
6836.36515 & $-1.52 \pm 0.13$ & $ -159$ & 102.4 &    997 & 4577--5055 & OPD (Lhires\,{\sc iii}) \\
6836.49685 & $-1.33 \pm 0.13$ & $ -149$ & 3.1 &    758 & 4505--4755 & CTIO \\
6837.38253 & $-1.40 \pm 0.13$ & $ -184$ & 100.0 &   1368 & 4577--5055 & OPD (Lhires\,{\sc iii}) \\
6838.51346 & $-1.65 \pm 0.13$ & $ -196$ & 3.7 &   1813 & 4505--4755 & CTIO \\
6839.46217 & $-1.86 \pm 0.13$ & $ -172$ & 100.0 &    395 & 4576--5058 & OPD (Lhires\,{\sc iii}) \\
6839.50001 & $-1.98 \pm 0.13$ & $ -169$ & 3.7 &   2037 & 4575--4800 & CTIO \\
6840.45062 & $-2.25 \pm 0.13$ & $ -149$ & 100.0 &    360 & 4576--5059 & OPD (Lhires\,{\sc iii}) \\
6840.46309 & $-2.53 \pm 0.13$ & $ -124$ & 91.3 &   1422 & 4575--4800 & SOAR \\
6840.49851 & $-2.43 \pm 0.13$ & $ -121$ & 3.7 &   2212 & 4575--5059 & CTIO \\
6841.39145 & $-2.08 \pm 0.35$ & $  -73$ & 100.2 &    350 & 3627--4939 & OPD (Lhires\,{\sc iii}) \\
6842.43257 & $-2.52 \pm 0.13$ & $ -136$ & 102.4 &   1203 & 4575--4800 & OPD (Lhires\,{\sc iii}) \\
6843.38235 & $-2.26 \pm 0.13$ & $ -183$ & 102.4 &    704 & 4574--5058 & OPD (Lhires\,{\sc iii}) \\
6844.41845 & $-2.07 \pm 0.13$ & $ -171$ & 100.3 &    616 & 4577--5055 & OPD (Lhires\,{\sc iii}) \\
6845.46145 & $-1.82 \pm 0.13$ & $ -267$ & 3.1 &    737 & 4577--5055 & CTIO \\
6847.45640 & $-1.62 \pm 0.13$ & $ -155$ & 3.7 &   1097 & 4577--5061 & CTIO \\
6847.47008 & $-1.98 \pm 0.13$ & $ -167$ & 91.2 &   2228 & 4505--4755 & SOAR \\
6850.38667 & $-1.87 \pm 0.13$ & $ -206$ & 102.4 &    658 & 4575--4800 & OPD (Lhires\,{\sc iii}) \\
6850.50978 & $-1.82 \pm 0.35$ & $ -182$ & 3.1 &    789 & 3628--4939 & CTIO \\
6850.96486 & $-1.93 \pm 0.13$ & $ -185$ & 72.2 &    747 & 4577--5055 & SASER \\
6851.38473 & $-2.30 \pm 0.13$ & $ -194$ & 102.4 &    875 & 4505--4755 & OPD (Lhires\,{\sc iii}) \\
6852.40542 & $-2.77 \pm 0.15$ & $ -217$ & 102.4 &    599 & 4574--4792 & OPD (Lhires\,{\sc iii}) \\
6852.49418 & $-2.93 \pm 0.13$ & $ -200$ & 25.9 &   1543 & 4577--5055 & CASLEO \\
6853.37790 & $-3.29 \pm 0.15$ & $ -171$ & 102.4 &    949 & 4614--5359 & OPD (Lhires\,{\sc iii}) \\
6853.49939 & $-2.79 \pm 0.13$ & $ -163$ & 25.7 &    329 & 4577--5055 & CASLEO \\
6854.80431 & $-2.57 \pm 0.30$ & $ -164$ & 3.2 &    732 & 4540--4760 & MJUO \\
6854.89023 & $-3.31 \pm 0.13$ & $ -176$ & 46.4 &    409 & 4577--5055 & SASER \\
6855.35972 & $-2.95 \pm 0.15$ & $ -229$ & 102.4 &    339 & 6396--6720 & OPD (Lhires\,{\sc iii}) \\
6855.45860 & $-3.09 \pm 0.30$ & $ -198$ & 3.1 &   1288 & 4540--4760 & CTIO \\
6855.47657 & $-2.79 \pm 0.15$ & $ -220$ & 25.8 &    667 & 4781--6681 & CASLEO \\
6855.77993 & $-2.69 \pm 0.15$ & $ -207$ & 3.2 &   1071 & 6397--6720 & MJUO \\
6855.86527 & $-3.10 \pm 0.47$ & $ -229$ & 56.5 &    785 & 4586--4818 & SASER \\
6855.94552 & $-2.74 \pm 0.15$ & $ -215$ & 206.1 &   1265 & 4570--4791 & SASER \\
6856.36061 & $-3.17 \pm 0.13$ & $ -263$ & 102.4 &    449 & 4577--5055 & OPD (Lhires\,{\sc iii}) \\
6856.45431 & $-3.16 \pm 0.15$ & $ -263$ & 3.7 &   2879 & 6397--6720 & CTIO \\
6856.46706 & $-3.51 \pm 0.13$ & $ -234$ & 91.4 &   1033 & 4505--4755 & SOAR \\
6856.47223 & $-2.35 \pm 0.30$ & $ -275$ & 25.8 &    356 & 4540--4760 & CASLEO \\
6856.87230 & $-3.25 \pm 0.47$ & $ -243$ & 52.4 &    409 & 4586--4818 & SASER \\
6857.45360 & $-3.23 \pm 0.15$ & $ -266$ & 3.1 &   1229 & 4570--4791 & CTIO \\
6858.40539 & $-3.27 \pm 0.15$ & $ -275$ & 102.4 &    622 & 4159--4913 & OPD (Lhires\,{\sc iii}) \\
6858.45238 & $-3.07 \pm 0.13$ & $ -271$ & 3.1 &   1329 & 4577--5055 & CTIO \\
6858.48475 & $-2.76 \pm 0.13$ & $ -300$ & 25.8 &    521 & 4575--4800 & CASLEO \\
6859.46549 & $-2.70 \pm 0.35$ & $ -272$ & 3.1 &    999 & 3625--4938 & CTIO \\
6859.49223 & $-3.03 \pm 0.30$ & $ -295$ & 25.7 &    777 & 4540--4760 & CASLEO \\
6860.38078 & $-2.73 \pm 0.15$ & $ -252$ & 102.4 &    802 & 4570--4791 & OPD (Lhires\,{\sc iii}) \\
6860.86021 & $-2.82 \pm 0.15$ & $ -251$ & 45.8 &    619 & 6020--6768 & SASER \\
6861.39604 & $-2.47 \pm 0.15$ & $ -309$ & 102.4 &    585 & 6019--6768 & OPD (Lhires\,{\sc iii}) \\
6862.40458 & $-2.65 \pm 0.13$ & $ -304$ & 102.4 &    575 & 4505--4755 & OPD (Lhires\,{\sc iii}) \\
6862.86946 & $-2.64 \pm 0.15$ & $ -296$ & 46.3 &     65 & 5736--6824 & SASER \\
6863.46562 & $-2.41 \pm 0.13$ & $ -326$ & 3.1 &   1162 & 4577--5055 & CTIO \\
6863.47455 & $-2.51 \pm 0.13$ & $ -304$ & 91.3 &   1498 & 4505--4755 & SOAR \\
6864.46694 & $-2.37 \pm 0.30$ & $ -323$ & 91.3 &   1031 & 4540--4760 & SOAR \\
6864.46759 & $-2.16 \pm 0.15$ & $ -335$ & 3.1 &   1180 & 4773--6676 & CTIO \\
6864.93373 & $-2.22 \pm 0.15$ & $ -326$ & 44.4 &   1271 & 6397--6720 & SASER \\
6865.87094 & $-1.69 \pm 0.15$ & $ -313$ & 44.1 &    430 & 6397--6720 & SASER \\
6866.51214 & $-0.93 \pm 0.13$ & $ -371$ & 3.0 &   1215 & 4505--4755 & CTIO \\
6867.41180 & $-0.88 \pm 0.30$ & $ -398$ & 100.3 &    501 & 4540--4760 & OPD (Lhires\,{\sc iii}) \\
6867.46441 & $-0.55 \pm 0.15$ & $ -396$ & 3.0 &    770 & 6397--6720 & CTIO \\
6867.46611 & $-0.69 \pm 0.13$ & $ -390$ & 91.3 &    670 & 4577--5055 & SOAR \\
6867.85881 & $-0.30 \pm 0.15$ & $ -368$ & 3.2 &    733 & 4578--4791 & MJUO \\
6868.39812 & $-0.44 \pm 0.13$ & $ -374$ & 100.2 &    379 & 4577--5055 & OPD (Lhires\,{\sc iii}) \\
6868.84061 & $+0.19 \pm 0.15$ & $ -681$ & 63.4 &     67 & 6397--6720 & SASER \\
6869.38458 & $-0.39 \pm 0.13$ & $ -336$ & 100.3 &    663 & 4577--5055 & OPD (Lhires\,{\sc iii}) \\
6869.46329 & $-0.28 \pm 0.15$ & $  -98$ & 91.2 &    331 & 6398--6720 & SOAR \\
6869.82462 & $-0.48 \pm 0.15$ & $ +530$ & 3.2 &    779 & 4578--4792 & MJUO \\
6869.85601 & $+0.35 \pm 0.13$ & $ +350$ & 63.1 &     85 & 4505--4755 & SASER \\
6869.94795 & $+0.07 \pm 0.35$ & $ -668$ & 208.6 &    267 & 3625--4938 & SASER \\
6870.38838 & $-0.24 \pm 0.35$ & $ -677$ & 100.2 &    590 & 3626--4938 & OPD (Lhires\,{\sc iii}) \\
6870.45625 & $+0.18 \pm 0.13$ & $ +530$ & 3.1 &    796 & 4505--4755 & CTIO \\
6870.46307 & $+0.21 \pm 0.15$ & $ +516$ & 91.5 &   1174 & 4570--4792 & SOAR \\
6871.42775 & $-0.03 \pm 0.15$ & $ +523$ & 100.3 &    489 & 4570--4788 & OPD (Lhires\,{\sc iii}) \\
6871.46362 & $+0.18 \pm 0.15$ & $ +530$ & 3.1 &    729 & 6355--6724 & CTIO \\
6871.49786 & $-0.11 \pm 0.13$ & $ -397$ & 26.0 &    391 & 4505--4755 & CASLEO \\
6871.50000 & $+0.09 \pm 0.13$ & $ -678$ & 39.5 &    503 & 4574--5058 & HST \\
6871.91627 & $+0.22 \pm 0.13$ & $ +528$ & 47.9 &    919 & 4505--4755 & SASER \\
6872.37893 & $-0.03 \pm 0.35$ & $ +531$ & 102.4 &    581 & 3627--4939 & OPD (Lhires\,{\sc iii}) \\
6872.46439 & $+0.07 \pm 0.15$ & $ -684$ & 3.1 &   1381 & 4586--4818 & CTIO \\
6872.85370 & $+0.64 \pm 0.13$ & $   +3$ & 60.8 &     79 & 4575--5059 & SASER \\
6872.93725 & $+0.31 \pm 0.15$ & $ +514$ & 207.5 &    379 & 6397--6720 & SASER \\
6873.38419 & $+0.02 \pm 0.15$ & $ +531$ & 102.4 &    592 & 4524--5100 & OPD (Lhires\,{\sc iii}) \\
6873.46368 & $+0.06 \pm 0.13$ & $ -684$ & 3.1 &   1000 & 4574--5058 & CTIO \\
6873.48259 & $+0.15 \pm 0.35$ & $ +523$ & 25.9 &    667 & 3628--4938 & CASLEO \\
6874.38414 & $+0.12 \pm 0.15$ & $ +531$ & 102.4 &    300 & 4586--4818 & OPD (Lhires\,{\sc iii}) \\
6874.45527 & $-0.04 \pm 0.15$ & $ +530$ & 3.1 &   1193 & 4526--5098 & CTIO \\
6874.47893 & $+0.14 \pm 0.15$ & $ +374$ & 27.0 &    845 & 4174--4925 & CASLEO \\
6874.85828 & $+0.17 \pm 0.15$ & $ +529$ & 72.1 &    118 & 6061--6804 & SASER \\
6874.93947 & $-0.25 \pm 0.13$ & $  -76$ & 207.7 &   3778 & 4574--5058 & SASER \\
6875.38311 & $+0.14 \pm 0.13$ & $ +531$ & 102.4 &    415 & 4505--4755 & OPD (Lhires\,{\sc iii}) \\
6875.47747 & $+0.13 \pm 0.35$ & $ +528$ & 26.0 &    687 & 3624--4939 & CASLEO \\
6876.38636 & $-0.00 \pm 0.15$ & $  -90$ & 102.4 &    446 & 6061--6803 & OPD (Lhires\,{\sc iii}) \\
6876.45922 & $-0.02 \pm 0.13$ & $ -231$ & 3.1 &    387 & 4574--5059 & CTIO \\
6876.48405 & $+0.03 \pm 0.13$ & $ +114$ & 26.0 &    836 & 4505--4755 & CASLEO \\
6877.40420 & $-0.18 \pm 0.03$ & $  -56$ & 102.4 &    615 & 4540--4760 & OPD (Lhires\,{\sc iii}) \\
6877.48745 & $+0.12 \pm 0.03$ & $ -674$ & 26.0 &    603 & 4560--4826 & CASLEO \\
6877.94054 & $-0.02 \pm 0.15$ & $ -212$ & 119.2 &    542 & 4572--4792 & SASER \\
6878.39833 & $-0.25 \pm 0.15$ & $  -90$ & 102.4 &    572 & 6060--6805 & OPD (Lhires\,{\sc iii}) \\
6878.48279 & $-0.39 \pm 0.15$ & $  -94$ & 3.1 &    758 & 6060--6806 & CTIO \\
6878.93662 & $+0.04 \pm 0.13$ & $ -104$ & 116.8 &    416 & 4577--5055 & SASER \\
6878.98294 & $-0.12 \pm 0.13$ & $  -27$ & 146.3 &     93 & 4505--4755 & SASER \\
6879.39884 & $-0.60 \pm 0.15$ & $  -67$ & 102.4 &    213 & 4561--5101 & OPD (Lhires\,{\sc iii}) \\
6879.46777 & $-0.45 \pm 0.15$ & $  -28$ & 3.1 &    969 & 4175--4925 & CTIO \\
6880.46780 & $-0.41 \pm 0.15$ & $  -60$ & 3.1 &    951 & 6060--6808 & CTIO \\
6880.48893 & $-0.49 \pm 0.13$ & $ -263$ & 102.4 &     86 & 4577--5055 & OPD (Lhires\,{\sc iii}) \\
6880.93207 & $+0.08 \pm 0.13$ & $  -32$ & 106.3 &    202 & 4505--4755 & SASER \\
6881.46594 & $-0.44 \pm 0.03$ & $  -48$ & 3.1 &   1034 & 4540--4760 & CTIO \\
6882.46242 & $-0.49 \pm 0.15$ & $  -55$ & 3.1 &   1248 & 5745--7657 & CTIO \\
6883.46622 & $-0.51 \pm 0.13$ & $  -28$ & 3.1 &    439 & 4577--5055 & CTIO \\
6883.47352 & $-0.38 \pm 0.13$ & $ -423$ & 27.0 &    887 & 4505--4755 & CASLEO \\
6883.83814 & $+0.29 \pm 0.03$ & $  -26$ & 3.2 &    600 & 4540--4760 & MJUO \\
6883.92037 & $-0.47 \pm 0.15$ & $ -131$ & 44.5 &    364 & 4488--5059 & SASER \\
6884.93750 & $-0.10 \pm 0.15$ & $ -158$ & 122.0 &    423 & 4167--4923 & SASER \\
6885.42504 & $-0.42 \pm 0.15$ & $  -79$ & 102.4 &     84 & 6065--6804 & OPD (Lhires\,{\sc iii}) \\
6885.46393 & $-0.46 \pm 0.13$ & $ -100$ & 3.1 &    806 & 4577--5055 & CTIO \\
6885.87352 & $-0.32 \pm 0.03$ & $  -72$ & 145.8 &    303 & 4540--4760 & SASER \\
6886.47066 & $-0.44 \pm 0.13$ & $  -38$ & 3.1 &    967 & 4577--5055 & CTIO \\
6887.46157 & $-0.47 \pm 0.13$ & $  +11$ & 3.7 &    755 & 4505--4755 & CTIO \\
6888.38076 & $-0.60 \pm 0.03$ & $  -33$ & 102.4 &    818 & 4540--4760 & OPD (Lhires\,{\sc iii}) \\
6888.93513 & $-0.43 \pm 0.13$ & $  -14$ & 115.4 &    168 & 4577--5055 & SASER \\
6889.40109 & $-0.47 \pm 0.03$ & $  -21$ & 102.4 &    597 & 4540--4760 & OPD (Lhires\,{\sc iii}) \\
6889.46464 & $-0.51 \pm 0.15$ & $ +527$ & 26.0 &    338 & 4165--4917 & CASLEO \\
6890.38792 & $-0.60 \pm 0.15$ & $  -10$ & 102.4 &    505 & 6054--6798 & OPD (Lhires\,{\sc iii}) \\
6890.84660 & $-0.70 \pm 0.13$ & $ +468$ & 59.0 &    140 & 4577--5055 & SASER \\
6891.37832 & $-0.68 \pm 0.13$ & $  +35$ & 102.4 &    643 & 4505--4755 & OPD (Lhires\,{\sc iii}) \\
6892.37992 & $-0.72 \pm 0.15$ & $  +35$ & 102.4 &   1030 & 4165--4917 & OPD (Lhires\,{\sc iii}) \\
6892.93822 & $-0.52 \pm 0.15$ & $  -32$ & 106.2 &    207 & 4628--5808 & SASER \\
6893.40410 & $-0.75 \pm 0.15$ & $  +12$ & 102.4 &    456 & 6054--6798 & OPD (Lhires\,{\sc iii}) \\
6894.39500 & $-0.89 \pm 0.13$ & $  -21$ & 102.4 &    380 & 4577--5055 & OPD (Lhires\,{\sc iii}) \\
6895.80517 & $-1.24 \pm 0.13$ & $  -22$ & 3.2 &    908 & 4505--4755 & MJUO \\
6895.81175 & $-0.60 \pm 0.13$ & $ -198$ & 71.8 &    218 & 4505--4755 & SASER \\
6895.93682 & $-0.54 \pm 0.13$ & $   +3$ & 95.0 &    112 & 4577--5055 & SASER \\
6896.79905 & $-1.18 \pm 0.15$ & $  -54$ & 3.2 &   1023 & 4163--4919 & MJUO \\
6896.82025 & $-1.07 \pm 0.13$ & $  -99$ & 71.1 &    296 & 4505--4755 & SASER \\
6897.79931 & $-1.11 \pm 0.13$ & $  -39$ & 3.2 &    810 & 4505--4755 & MJUO \\
6898.80025 & $-1.08 \pm 0.13$ & $  -30$ & 3.2 &    955 & 4505--4755 & MJUO \\
6899.41741 & $-1.22 \pm 0.22$ & $  -56$ & 102.4 &    248 & 4540--4760 & OPD (Lhires\,{\sc iii}) \\
6899.79987 & $-1.36 \pm 0.03$ & $  -20$ & 3.2 &    931 & 4591--4817 & MJUO \\
6900.39932 & $-1.56 \pm 0.15$ & $  -44$ & 102.4 &    262 & 4581--4792 & OPD (Lhires\,{\sc iii}) \\
6900.80626 & $-1.64 \pm 0.15$ & $  -23$ & 3.2 &   1231 & 4163--4919 & MJUO \\
6901.80112 & $-1.71 \pm 0.15$ & $   -9$ & 3.2 &    956 & 6054--6798 & MJUO \\
6902.37996 & $-2.14 \pm 0.13$ & $  -56$ & 102.4 &    540 & 4577--5055 & OPD (Lhires\,{\sc iii}) \\
6902.79568 & $-1.64 \pm 0.13$ & $   -3$ & 3.2 &    750 & 4505--4755 & MJUO \\
6902.92365 & $-2.02 \pm 0.15$ & $  +30$ & 70.9 &    169 & 4573--5755 & SASER \\
6903.39135 & $-2.03 \pm 0.13$ & $  -33$ & 102.4 &    410 & 4505--4755 & OPD (Lhires\,{\sc iii}) \\
6903.80951 & $-2.03 \pm 0.13$ & $  -21$ & 3.2 &   1365 & 4575--4800 & MJUO \\
6904.80022 & $-1.68 \pm 0.13$ & $  -33$ & 3.2 &    920 & 4577--5055 & MJUO \\
6905.40209 & $-1.59 \pm 0.15$ & $  -79$ & 102.4 &    253 & 4163--4919 & OPD (Lhires\,{\sc iii}) \\
6905.81433 & $-1.42 \pm 0.13$ & $  -25$ & 3.2 &    605 & 4577--5055 & MJUO \\
6905.94347 & $-1.15 \pm 0.22$ & $  -14$ & 100.8 &    351 & 4540--4760 & SASER \\
6906.38324 & $-1.51 \pm 0.13$ & $  -21$ & 102.4 &    419 & 4577--5055 & OPD (Lhires\,{\sc iii}) \\
6906.80165 & $-1.58 \pm 0.15$ & $  -23$ & 3.2 &    864 & 4489--5031 & MJUO \\
6906.83424 & $-1.46 \pm 0.13$ & $  -56$ & 102.4 &    494 & 4577--5055 & OPD (Lhires\,{\sc iii}) \\
6907.37853 & $-1.45 \pm 0.13$ & $  -79$ & 102.4 &    317 & 4577--5055 & OPD (Lhires\,{\sc iii}) \\
6907.80059 & $-1.35 \pm 0.15$ & $  -18$ & 3.2 &    846 & 4163--4919 & MJUO \\
6907.80869 & $-1.02 \pm 0.13$ & $  -52$ & 68.9 &    282 & 4577--5055 & SASER \\
6908.38139 & $-1.26 \pm 0.13$ & $  -44$ & 102.4 &    159 & 4577--5055 & OPD (Lhires\,{\sc iii}) \\
6908.87070 & $-1.08 \pm 0.47$ & $  -11$ & 3.2 &    457 & 4586--4818 & MJUO \\
6909.38647 & $-1.05 \pm 0.15$ & $  -33$ & 102.4 &    272 & 4619--5199 & OPD (Lhires\,{\sc iii}) \\
6910.38143 & $-1.04 \pm 0.15$ & $  -67$ & 102.4 &    279 & 4163--4919 & OPD (Lhires\,{\sc iii}) \\
6910.80338 & $-0.76 \pm 0.47$ & $  -57$ & 70.6 &    470 & 4586--4817 & SASER \\
6910.82866 & $-0.86 \pm 0.15$ & $  -67$ & 102.4 &   1054 & 4622--5195 & OPD (Lhires\,{\sc iii}) \\
6911.30601 & $-0.77 \pm 0.47$ & $  -60$ & 398.8 &     79 & 4586--4818 & SASER \\
6911.38002 & $-1.07 \pm 0.47$ & $  -79$ & 102.4 &    378 & 4586--4818 & OPD (Lhires\,{\sc iii}) \\
6911.80060 & $-0.71 \pm 0.13$ & $  -90$ & 102.4 &    652 & 4577--5055 & OPD (Lhires\,{\sc iii}) \\
6911.95672 & $-0.83 \pm 0.47$ & $ -123$ & 94.5 &    314 & 4586--4818 & SASER \\
6912.94395 & $-0.72 \pm 0.13$ & $ -180$ & 89.7 &    226 & 4577--5055 & SASER \\
6914.94813 & $-0.99 \pm 0.47$ & $ -145$ & 100.2 &    421 & 4586--4818 & SASER \\
6915.24671 & $-0.90 \pm 0.47$ & $ -126$ & 473.3 &    124 & 4586--4818 & SASER \\
6915.94736 & $-1.17 \pm 0.13$ & $ -128$ & 98.0 &    792 & 4577--5055 & SASER \\
6916.94919 & $-0.71 \pm 0.47$ & $  -57$ & 117.3 &    114 & 4586--4818 & SASER \\
6917.78979 & $-0.78 \pm 0.15$ & $ -102$ & 102.4 &    645 & 4618--5190 & OPD (Lhires\,{\sc iii}) \\
6919.26677 & $-0.67 \pm 0.13$ & $  -85$ & 453.1 &    477 & 4577--5055 & SASER \\
6923.24250 & $-0.31 \pm 0.47$ & $  -81$ & 54.3 &     77 & 4586--4817 & SASER \\
6924.23432 & $-0.67 \pm 0.47$ & $ -192$ & 137.9 &     84 & 4586--4817 & SASER \\
6924.23432 & $-0.65 \pm 0.13$ & $  -95$ & 70.4 &     97 & 4577--5055 & SASER \\
6928.50000 & $-0.40 \pm 0.47$ & $  -37$ & 39.5 &    462 & 4589--4818 & HST \\
6938.83294 & $-0.24 \pm 0.15$ & $ -102$ & 102.4 &    708 & 4163--4919 & OPD (Lhires\,{\sc iii}) \\
6941.90973 & $-0.02 \pm 0.13$ & $ -643$ & 3.0 &    157 & 4577--5055 & CTIO \\
6942.91028 & $+0.04 \pm 0.47$ & $  -90$ & 3.1 &    213 & 4586--4818 & CTIO \\
6943.90679 & $-0.09 \pm 0.13$ & $  -74$ & 3.1 &    595 & 4577--5055 & CTIO \\
6949.24499 & $-0.27 \pm 0.13$ & $  -81$ & 36.3 &    121 & 4577--5055 & SASER \\
6950.87840 & $+0.01 \pm 0.47$ & $ -179$ & 3.1 &   1301 & 4586--4818 & CTIO \\
6951.87621 & $+0.03 \pm 0.15$ & $ -114$ & 3.1 &   1268 & 4598--5169 & CTIO \\
6952.86748 & $-0.03 \pm 0.13$ & $  -81$ & 3.1 &    725 & 4577--5055 & CTIO \\
6953.88255 & $-0.04 \pm 0.47$ & $   +3$ & 3.1 &   1128 & 4586--4818 & CTIO \\
6954.83416 & $-0.06 \pm 0.13$ & $  -13$ & 3.1 &    796 & 4577--5055 & CTIO \\
6955.86423 & $-0.03 \pm 0.13$ & $ +223$ & 3.1 &    982 & 4577--5055 & CTIO \\
6956.87673 & $+0.03 \pm 0.15$ & $  -27$ & 3.1 &   1281 & 4595--5168 & CTIO \\
6957.89313 & $-0.01 \pm 0.13$ & $   +3$ & 3.1 &    914 & 4577--5055 & CTIO \\
6958.86680 & $-0.07 \pm 0.15$ & $  +84$ & 3.1 &    850 & 4282--5799 & CTIO \\
6959.84998 & $+0.05 \pm 0.13$ & $  -44$ & 3.1 &    951 & 4577--5055 & CTIO \\
6960.73148 & $-0.00 \pm 0.13$ & $  -44$ & 102.4 &    599 & 4577--5055 & OPD (Lhires\,{\sc iii}) \\
6961.85099 & $-0.00 \pm 0.15$ & $  +33$ & 3.1 &   1009 & 4164--4918 & CTIO \\
6964.86782 & $+0.02 \pm 0.15$ & $   -3$ & 3.1 &    904 & 4164--4918 & CTIO \\
6965.86504 & $+0.03 \pm 0.15$ & $   -5$ & 3.1 &   1244 & 4168--4915 & CTIO \\

\enddata

\tablenotetext{a}{Resolution element.}
\tablenotetext{b}{Signal-to-noise ratio per resolution element.}

\end{deluxetable*}

\end{document}